\documentclass[rmp,reprint,nofootinbib,aps,superscriptaddress]{revtex4-1}
 \pdfoutput=1

\usepackage{epsfig}
\usepackage{longtable}
\usepackage[T1]{fontenc}
\usepackage[latin9]{inputenc}
\usepackage{graphicx}
\usepackage{amsmath}
\usepackage{rotating}           
\usepackage{color}
\usepackage{ulem}
\usepackage{pdfsync}
\usepackage{fancyvrb}

\makeatletter

\newcommand{\fig}[1]{Figure (\ref{#1})}
\newcommand{\tab}[1]{Table (\ref{#1})}
\newcommand{\eq}[1]{Eq. (\ref{#1})}

\newcommand{\Ref}[1]{Ref.\cite{#1}}
\newcommand{\se}[1]{Sec. (\ref{#1})}

\newcommand{\la}[1]{ \label{#1}}
\renewcommand{\a}{\alpha}
\renewcommand{\b}{\beta}
\renewcommand{\d}{\delta}

\newcommand{\h}{\mathbf{H}}
\renewcommand{\r}{\mathbf{r}}
\newcommand{\s}{\sigma}
\renewcommand{\O}{\mathcal{O}}

\newcommand{\bsubs}{\begin{subequations}}
\newcommand{\esubs}{\end{subequations}}

\providecommand{\ba}{\begin{align}}
\providecommand{\ea}{\end{align}}
\renewcommand{\H}{\mathcal{H}}
\usepackage{multirow}
\newcommand{\be}{\begin{equation}}
\newcommand{\ee}{\end{equation}}

\newcommand{\bea}{\begin{eqnarray}}
\newcommand{\eea}{\end{eqnarray}}

\begin{document}

\title{Real Space Renormalization in Statistical Mechanics}
\author{Efi Efrati}
\email{efrati@uchicago.edu}
\affiliation{James Franck Institute, The University of Chicago.
929 E. 57 st, Chicago, IL 60637, USA}
\author{Zhe Wang}
\affiliation{James Franck Institute, The University of Chicago.
929 E. 57 st, Chicago, IL 60637, USA}
\author{Amy Kolan}
\affiliation{James Franck Institute, The University of Chicago.
929 E. 57 st, Chicago, IL 60637, USA}
\affiliation{St. Olaf College, Northfield, MN,USA}
\author{Leo P. Kadanoff}
\affiliation{James Franck Institute, The university of Chicago.
929 E. 57 st, Chicago, IL 60637, USA}
\affiliation{The Perimeter Institute, Waterloo, Ontario, Canada}

\begin{abstract}
This paper discusses methods for the construction of approximate real space renormalization transformations in
statistical mechanics.  In particular, it compares two methods of transformation: the ``potential-moving'' approach most used in the period 1975-1980 and the ``rewiring method'' as it has been developed in the last five years. These methods both employ a parameter, called $\chi$ or $D$ in the recent literature, that measures the complexity of the localized stochastic variable forming the basis of the analysis.   Both methods are here exemplified by calculations in terms of fixed points for the smallest possible values of $\chi$.  These calculations describe three models for two-dimensional systems: The Ising model solved by Onsager, the tricritical point of that model, and the three-state Potts model.

The older method, often described as  {\em lower bound renormalization theory},  provides a heuristic method giving reasonably accurate results for critical indices at the lowest degree of complexity, i.e. $\chi=2$.  In contrast, the {\em rewiring} method, employing ``singular value decomposition'', does not perform as well for low $\chi$ values but offers an error that apparently decreases slowly toward zero as $\chi$ is increased. It appears likely that no such improvement occurs in the older approach.

A detailed comparison of the two methods is performed, with a particular eye to describing the reasons why they are so different.  For example, the older method is based on the analysis of spins, simple stochastic variables located at lattice sites. The new method uses ``indices'' describing linear combinations of different localized configurations.  The old method quite naturally employed fixed points for its analysis; these are hard to use in the newer approach.  A discussion is given of why the fixed point approach proves to be hard in this context.

In the new approach the calculated the thermal critical indices are satisfactory for the smallest values of $\chi$ but hardly improve as $\chi$ is increased, while the magnetic critical indices do not agree well with the known theoretical values.
\end{abstract}
\maketitle

\tableofcontents

\section{Introduction}
\subsection{History: the conceptual foundations}
The renormalization group \cite{Cardy,LPK00,MarisKad,Nelson},  provides a theoretical understanding of singular problems in
statistical mechanics \cite{Domb}, particularly ones involving phase transitions.  There are two main branches of analysis
based upon this method, one involving work in momentum or wave vector space \cite{Wilson-Kogut}, the other involving so-
called ``real-space'' methods.    In this review, we follow the latter approach.

Both approaches make extensive use of the following conceptual ideas, {\em vis}
\begin{itemize}
\item {Scale invariance. }   Singularities in statistical mechanics tend to be connected with behaviors that are the same at different length scales.   Critical points of phase transitions have correlations at all length scales.
\item {Scale covariance. }   Near the phase transition, many physical quantities vary as  powers of characteristic lengths that
describe the system, or of lengths describing the quantities themselves,  or as powers of ``fields'' that measure the
deviation of thermodynamic quantities from criticality.  These powers characterize the phase transition.  They are called {\em critical indices.}
\item {Fixed point.  }The scale-invariance is described by a Hamiltonian or free-energy-function that has elements that are independent of length scale.  As a result, one might expect that, for example, the Hamiltonian or the free
energy function that describes the system will not change when the length scale changes.  This unchanging behavior is
described as ``being at a fixed point''.
\item {Renormalization. } A transformation that describes the results of changes in the length scale.  Usually this transformation will not change the Hamiltonian or free energy describing the fixed point.   That is the reason for the name, {\em fixed point}.
\item {Universality.}   Near the phase transition, many different physical systems show identical behavior of the quantities
that describe critical behavior. Since these quantities are descriptive of scale-invariant behavior these descriptive quantities all can be seen at large length scales.
\item {Universality classes.}  There are many critical points with a wide variety of different origins.  Nonetheless these fall into relatively few {\em universality classes}, each class being fully descriptive of all the details of a given critical behavior.
\end{itemize}
The  behavior of different critical systems can be, in large measure, classified
by describing the dimension and other topological features of the system, and then describing some underlying symmetry that
plays a major role at the critical point.  Thus, the model that Lars Onsager  solved, the Ising model, \cite{Onsager,Ising, Lenz} is mostly described by saying it is a two-dimensional system with a spin at each point. The spin can point in one of two directions. The model has a symmetry under flipping the sign of a spin, so that it can describe a magnetic phase transition.  It is equally well descriptive of a two-dimensional liquid in which the basic symmetry is in the interchange of high density regions with low density ones, so that it describes a liquid-gas phase transition.  Any model with the appropriate
symmetry and dimensionality and the right range of interaction strengths is likely to describe both situations, and many
others.  The Ising model constitutes the simplest model of this kind.

\subsection{Statistical Variables}
There are many models and real systems that exhibit critical behavior \cite{Ma,Stanley,67Review,NBS}.  All of those with short-range interactions and spatial homogeneity have the same kind of characteristic behavior.  One starts from a statistical ensemble, that is a very large system of stochastic variables, called $\{\s_\r\}$, where $\r$ defines a position in space.  The statistical calculation is defined by probability distribution,  given as an expression of the form $\exp{(-\b \h\{\s_\r\})}$, where $\b$ is the inverse temperature and $\h$ is the Hamiltonian for the statistical system.  One then uses a sum over all the stochastic variables, defined by the linear operation denoted as {\em trace}, to define a thermodynamic quantity the free
energy, $F$, as
\bsubs
\be
e^{-\b F} =\text{Tr}_{\{\s_\r\}}~\left(\,e^{-\b \h\{\s_\r\}}\,\right).
\la{F}
\ee
\eq{F} gives the problem formulation for statistical physics introduced by Boltzmann and Gibbs and directly used for renormalization calculations through the 1980s.
In recent years, a slightly different formulation has taken hold. Since the Hamiltonian is most often a sum of terms, each containing a few spatially-neighboring $\s_\r$ values, one can write the free energy as a sum of products of blocks:
\be
e^{-\b F}=\text{trace}_{\{\s_\r\}}\prod_\mathbf{R} \text{BLOCK}_\mathbf{R},
\ee
each block depending on a few statistical variables.  This formulation applies equally well to the older and the new formulations of the statistical mechanics.  Lately, statistical scientists have realized the advantage of a
particular special form of writing the product of blocks,  called the {\em tensor network representation}.  In this representation, similarly to vertex models, the statistical Boltzman weights are associated with vertices (rather than bonds) \cite{Baxter}.

The {\em tensor network representation} describes the connectivity and interdependence of blocks and statistical variables. Because of locality the numerical value each block attains depends on a small  number of statistical variables. Every statistical variable in turn affects the numerical values  in a small number of different blocks. This allows the identification of a statistical variable assuming $\chi$ different values with an {\em index} assuming the values $\{1 ,2, \cdots,\chi\}$. The blocks are linked because each index appears in precisely two blocks. The blocks then reduce to tensors whose rank is determined by how many different indices determine the values assumed by a given block. Every configuration corresponds to a specific choice of indices. It is believed, but not proven, that this kind of representation forms a link to the fundamental description of the statistical problem \cite{Vidal0}. The free energy calculation which follows by summation over all possible configurations of the statistical variables reduces to a tensor product tracing out all the mutual index values.
 \be
e^{-\b F} =\text{Tr}_{i,j,k,\cdots}~\prod T_{ijkl}
\la{FT}
\ee
\esubs

The tensor {\em indices} are indirect representations of the original statistical variables. Each value of a given index may represent a sum, with coefficients that can be positive or negative, of the weights of statistical configurations in the system. Moreover,  this representation permits a kind of gauge invariance for each index at each point in space, giving the index variable a new meaning.  For example, if the index $i$ appears in two tensors $T^1_{ijkl}$ and $T^2_{ipqr}$, then the index transformation,
\be
T^1_{ijkl}\to   \sum_m  O_{i,m} T^1_{mjkl},\quad
T^2_{ipqr}\to   \sum_m  O_{i,m} T^2_{mpqr},\quad
\ee
for $\sum_m  O_{i,m} O_{j,m} =\delta_{i,j}$, will leave the partition function unchanged.   This important formal property underpins the newer statistical calculations.

\subsection{Renormalization}
The basic theory describing this kind of behavior was derived by Wilson \cite
{Wilson}, based in part upon ideas derived earlier \cite{Gell-MannLow, Widom, PP, LPK1966,LPK09,LPK10,LPK11}. The first element of the theory is the concept of a {\em  renormalization transformation.}   This is a change in the description of an ensemble of statistical
systems, obtained by changing the length scale upon which the system is described. Such a transformation may be applied to any statistical system, including ones which are or are not at a critical point.  There is a whole collection of methods for constructing such renormalization transformations and describing their properties.  This paper will be concerned with describing one class of such transformations, the {\em real-space} transformations. These are ones that employ the description of the ensemble in ordinary space (or sometimes space-time) to construct a description of the renormalization process.

The ensemble is parameterized by a set of coupling constants, ${\bf K}= \{K_j\}$. These couplings might describe the spin interactions of the early renormalization  schemes,  with the subscripts denoting couplings to different combinations of spin operators. Alternatively the $K$'s may be parameters that determine the  tensors.
The renormalization transformation increases some characteristic distance describing the system,
usually the distance between neighboring lattice points on a lattice defining the spatial structure, so that this distance changes according to $ a' =\d L ~a$.   Correspondingly, the renormalization transformation changes the coupling parameters to new values which we denote by $\bf{K}'$.  These new couplings depend upon the values of the old ones, so that
\be
{\bf K}' =\mathcal{R}( \bf{K})
\la{renorm}
\ee
Here, the function $\mathcal{R}$ represents the effect of the renormalization transformation.

\subsection{Fixed Point}
The renormalization theory is particularly powerful at the critical point. This application of the
theory is  based upon the concept of a {\em fixed point}, an ensemble of statistical systems that describe the behavior of all individual statistical systems within a particular universality class.    Since the critical point is itself invariant under scaling transformations, the ensemble in question is invariant under a renormalization transformation.  It is said to be at a {\em fixed point}.  The fixed point is represented by a special set of couplings, $\bf{K}^*$, that are invariant under the renormalization transformation
\be
\bf{K}^* =\mathcal{R}( \bf{K}^*)
\la{renorm}
\ee

\subsection{Critical indices}
The most important physical effects are obtained by studying the behavior of the renormalization transformation in the vicinity of the critical points. This behavior is in turn best described by a set of critical indices describing the scaling of the singular part of measurable physical quantities near the critical point.
One writes  the deviation of the Hamiltonian from its fixed point value as
\be
-\b \h =-\b \h^* +\sum_\a h_\a~ S^\a
\la{}
\ee
where the components $h_\a$ represent the small deviations of the coupling constant $K_\a$ from their fixed point values, while $S^\a$ are the extensive stochastic operator conjugates to the  $K_\a$.  The free energy undergoes a change in value produced by these variations of the form:
\be
- \b \d F = \sum_\a C_\a  \eta_\a L^{d-x^\a}
\la{deviation}
\ee
Here, the $ \eta_\a $ describes linear combinations of the $h_j$, each $ \eta_\a $  defining a different kind of covariant {\em scaling operator} that describes a particular type of scaling near the critical point. In the usual critical phenomena problems one such operator describes the field thermodynamically conjugate to the order parameter whose symmetry breaking produces the phase transition, and another such field, conjugate to the energy density, is the deviation of temperature from its critical value.  Still other operators, such as the stress tensors, play  important roles in the critical behavior, but have symmetries that prevent them from appearing at first order in this expansion.

In \eq{deviation},  $L$ is the linear dimension of the system, $d$ is the dimensionality of the system,  while $C_\a$ is a relatively unimportant expansion coefficient.  The crucial quantity in this equation is $x^\a$, the {\em critical index} defining the scaling properties of the scaling operator.
For the usual always-finite scaling operators the exponents $x^\a$ are positive. Operators for which the corresponding critical exponents lie between zero and the dimension of the system, $d$, are called {\em relevant operators}. These play a major  role in the thermodynamics. Operators for which the corresponding critical exponents are greater than $d$ are called {\em irrelevant operators} and do not contribute to the to the singular behavior of the thermodynamic functions \cite{KW71,Weg76a}.
In our work below, we shall compare the values of the relevant $x$'s as they emerge from the approximate numerical
renormalization theory with the exact values that are often known from other numerical work or exact theories \cite
{BPZ, FMS}.

\subsection{Response analysis}\la{Response}
The behavior of the renormalization transformation in the vicinity of the critical points is quantified by the response matrix relating small changes in the couplings, $K_j$, with the small changes they induce in the renormalized couplings , $K_i'$.
\be
{B_{i}}^j= \left. \frac{dK'_i}{dK_j}\right| _{K=K^*}.
\ee This matrix has right and left eigenfunctions defined as
\be
\sum_j {B_{i}}^j {\psi_j}^\a =E^\a {\psi_i}^\a    \text{~~~and~~~} \sum_i {\phi_\a}^i {B_{i}}^j  =E^\a {\phi_\a}^j
\la{eigens}  \ee
 The eigenvalues define the scaling properties of the exact solution.  The  operators' scaling is
defined by \eq{deviation}.   The eigenvalues directly determine this scaling since  \be
E^\a =  (\delta L)^{y^\a}=(\delta L)^{d-x^\a}
\la{y} \ee
with $\delta L$ being the change in length scale produced by the renormalization.
\footnote{If $\d L_1$ and $\d L_2$ are two rescaling factors, then the form of the eigenvectors as a function of these rescaling factors satisfy
$E(\d L_1 \cdot\d L_2)=E(\d L_1)E(\d L_2)$. It follows that $E(1)=1$ and that $E'(\d L_1)/E(\d L_1) = C$ for some constant $C$, leading to
$E(\d L)\propto \d L^{y}$ \cite{Gol92}.}
Different authors describe their results in terms of $E_\a$, or $y^\a$, or $x^\a$.  In this paper, we use the last descriptor.

The eigenfunctions $\phi$ and $\psi$  can be used to construct a set of densities, $o^\a(\r)$, of operators called scaling operators since they have simple properties under scale transformations.  The combinations
\bsubs
\be
\sum_j   s^j(\r) {\psi_j}^\a = o^\a(\r)
\la{define-o} \ee
define $o^\a(\r)$ as the densities for the {\em scaling operator}. These operators and their extensive counterparts $O^
\a=\sum_\r o^\a(\r)$ respectively scale like distances to the
power $-x^\a$ and $y^\a$ respectively.  The other coefficient in the eigenvalue analysis, $\phi^a_i$ can be interpreted by saying that
$s^i$ generates a combination of fundamental operators according to
\be
s^i(\r) =\sum_\a {\phi_\a}^i o^\a(\r)
\la{expand-s} \ee
\esubs
To make \eq{define-o} and \eq{expand-s} work together, we must define the eigenvectors so that they are normalized and
complete
\be
\la{orthonormal}
\sum_i {\phi_\a}^i {\psi_i}^\b  = {\delta_{\a}}^\b  \text{~~~and~~~}   \sum_\a {\phi_\a}^j {\psi_i}^\a  = {\delta_{i}}^j
\ee

\subsection{Requirements on approximations}

The concepts of renormalization and scale invariance lead naturally to the identification of scaling and universality and have contributed to the fundamental understanding of critical phenomena. There is also a more practical aspect of the renormalization concepts that allows one to predict the location of phase transitions of specific systems and describe their nature in terms of the critical exponents. However, for most systems, carrying out the actual renormalization cannot be done exactly.  Instead, some approximation method must be used to find an  approximate renormalization transformation. We hope that the approximation method might give an informative picture of the physical system, that it might be numerically accurate, and that it might be improvable so that more work can lead to better results.

\subsection{History of real-space methods}
The first heuristic definition of a real space renormalization was given in  \cite{LPK1966}.  After Wilson and Fisher \cite{Wilson-Kogut,Wilson-Fisher} demonstrated the viability of the renormalization approach by inventing the $\epsilon$ expansion, Neimeijer and Van Leeuwen \cite{N-van L,N-van L2} described a method for doing a numerical calculation of the renormalization function, $R$, in terms of a small number of different couplings.  These methods were then described in one dimension \cite{KN} and applied \cite{KH}.    From the point of view of this paper, an important advance occurred when a variational method was invented \cite{K75} and extensively employed \cite{KH,Bu,Bu2,Bu3,Bu4,Bu5,Bu6,Knops,Nijs-Knops}. This method was described as a {\em lower bound calculation} since it permitted calculations that gave a lower bound on possible values of the free energy. This approach permitted reasonably accurate and extensive calculations of critical
properties in two and higher dimensions.   However, as the 1970s came to an end, the lower bound method fell into disuse.

In part, the disuse arose because interest turned from problems in classical statistical mechanics to quantum problems. Although path-integral methods permit one to convert a quantum problem into one in classical statistical mechanics, the lower bound method seemed to work best when the problem had the full rotational symmetry of its lattice, and hence did not apply to many quantum problems.  In 1992 White \cite{White92} invented a quantum mechanical real space renormalization scheme that worked beautifully for finding the properties of one dimensional quantum systems via numerical analysis. This success started a large school of work aimed at these problems and analogous problems in higher dimensions \cite{Zhao,Scholl, V,EV}.

White's method looked very different from the real-space work of the 1970's. It did, however, have an important provenance in Wilson's numerical solution of the Kondo problem \cite{WilsonKondo}. White studied the approximate eigenstates involving long chains of correlated spins, and how those long chains interacted with small blocks of spins.  There was also earlier work  that focused on blocks of a small number of spins.  In the course of time, connections among the different approaches began to be appreciated.   As pointed out by, for example, Cirac and Verstraete \cite{CiracV}, the correlations within wave functions were produced by summing products of correlations on small blocks, producing situations described as ``tensor product states''.   Levin and Nave \cite{LevinNave},  Gu and Wen\cite{GW}, and Vidal and coworkers\cite{V,EV} described how an accurate analysis could be constructed based on the correlations among statistical variables located at a very small number of nearby lattice sites.

On the one-dimensional lattice, Vidal \cite{V} and coworkers use two or three neighboring lattice sites as the basis of the correlations.  In higher dimensions Levin and Nave use a hexagonal construction  in which the basic variables are three tensor indices, each independently taking on integer values grouped around a three-legged lattice site.  Gu and Wen correspondingly use a four index tensor describing the intersections in a square lattice (see \fig{BasicTensor}) .

 \begin{figure}[h]
\begin{center}
\includegraphics[height=6cm ]{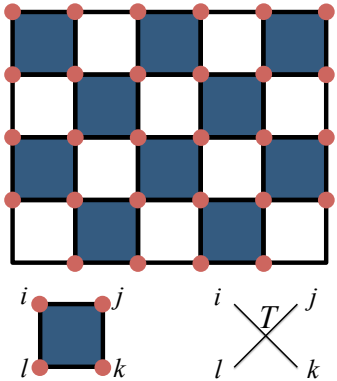}
\end{center}
\caption{The basic tensor network used here for the SVD renormalization calculations. Tensors are represented by blue solid colored squares. Red circles denote the position of the tensor's indices. Every tensor has four indices. Every index assume integer values between one and $\chi$, and is shared between exactly two tensors. Four indices determine the configuration of the statistical variable, and the corresponding tensor entry gives the statistical weight of the configuration. Note that the interactions represented by the tensors occupy half the available space.  The left inset shows the labeling of the indices; the right inset shows the same tensor in the stick-figure usually used in the literature.}
\la{BasicTensor}
\end{figure}

\subsection{Comparisons}
The main points of difference  between the work of the 1970s and that of the last two decades include
\subsubsection{Stochastic variables}\la{Variables}
We have already mentioned  that the 1975 scheme uses spins while the recent scheme employs much more complex spatial structures labeled by tensor indices. Both approaches need to reflect the underlying symmetry of the problem at hand, for example the spin flip symmetry of an Ising model.  The early work used spin variables that directly reflected the  symmetry. \footnote{However, there were occasional uses of more complex variables.  In Burkhardt's\cite{Bu} Ising model calculation, the ``spin'' variable could take on three values: $\pm 1$ and zero.  The last value reflecting a hole unoccupied by a magnetic spin.}
 In contrast, the more recent work has replaced summation over spin variables by sums over  tensor indices.   The basic symmetries are  hidden in the structure of these  tensors.  In using this tensor representation, recent workers have used universality to say that they can use any problem-definition that reflects a desired symmetry. They then also argue that the proper meaning of the tensor indices will give them direct access to the deep structure of the statistical mechanics problem \cite{SW}.

Each tensor index can take on $\chi$ possible values, representing that number of different configurations of the system.   Recent workers  believe, but have not proven, that they can get perfect accuracy
when $\chi$ is infinite. Consequently they reach for approximation methods that permit them to increase $\chi$ until it reaches quite large values. ( Note that these indices with their large number of possible values can simultaneously approximately
represent many kinds of different variables:  many-component vectors, Ising spins, or continuous variables. )   In contrast the earlier workers felt that arbitrary accuracy would not be available to them.  The best that was expected was a qualitatively accurate description of the problem.

We use the term {\em summation variables} to describe both the spins of the earlier work, and the stochastic variables linked to the tensor indices more recently used.

\subsubsection{Geometric structure}
Another difference can be seen in the geometric structures used to describe the interactions among the summation
variables.   In the tensor work the summation variables, in a similar fashion to their role in vertex models, (e.g. see \cite{Baxter}), are associated with bonds and their interactions are associated with vertices. This constrains each summation variable to participate in exactly two interactions (connect two vertices). The interactions, however, are less constrained and typically group together several summation variables around a rank $m$ vertex.

In the earlier renormalization work, in contrast, the summation variables are associated with vertices and thus may participate in more than two interactions. The interactions are associated with blocks of summation variables allowing more than only pairwise interactions.

This difference not only manifests in the formulation of the partition function of each of the representations but more importantly restricts the placement of the rescaled summation variables and their interactions. In the earlier work new summation variables could be placed arbitrarily provided the interactions they participate in can be formulated in term of the old interaction blocks. In the tensor representation the binary interaction structure must be preserved when introducing new summation variables. Thus every introduction of a new summation variable is necessarily associated with changing the interaction connectivity of the old variables.

\subsubsection{Calculational strategy}
The earlier work found the properties of critical points via a method based upon the analysis of fixed points.  First, the critical system was brought to a fixed point.  Critical indices were then calculated by looking at the growth or decay under renormalization of small perturbations  about the fixed point, using a method based upon eigenvalues (see \se{Response}.)  The main output of the calculation were a set of critical indices which could be compared among calculations and with theoretical results.

In contrast, tensor analysts seldom calculate fixed points.\footnote{Notable exceptions include the Hamiltonian work of Vidal and coworkers \cite{EV,V}, in which a fixed point Hamiltonian is indeed calculated.  For statistical rather than quantum problems, fixed point studies were done by \Ref{Aoki} and \Ref{HB}. This fixed point analyses, however, were only be carried out for small values of $\chi$.} Instead they calculate free energies and other thermodynamic quantities by  going through a large number of renormalizations,  usually increasing the value of $\chi$ as they go.  (As we shall see, it is natural to square the value of $\chi$ in each tensor renormalization.)  When they reach a maximum convenient value of $\chi$ they employ approximations that enable them to continue to renormalize with fixed $\chi$.   These calculations then show  the thermodynamic behavior near criticality.

The non-appearance of fixed points in many of the tensor calculations provides an important stylistic contrast between that work and the studies of the 1975-era.    The calculation of fixed points for the critical phenomena problems permits the direct calculation of critical indices and thus offers many insights into the physics of the problem. The insights are obtained by keeping track of and understanding every coupling constant used in the analysis.  This is easy when there are, as in  \Ref{K75}, sixteen couplings.  However, the more recent tensor-style work often employs indices which are summed over hundreds of values, each representing a sum of configurations of  multiple spin-like variables. All these indices are generated and picked by the computer.  The analyst does not and cannot keep track of the meaning of all these variables. Therefore, even if a fixed point were generated, it would not be very meaningful to the analyst. In fact, the literature does not seem to contain much information about the values and consequences of fixed points for the new style of renormalization.

The fixed point method seems more fundamental and preferable, but offers major challenges when the value of $\chi$ is large.

\subsection{Plan of paper}

The next section describes the block spin and the rewiring methods employing singular value decomposition (SVD) used for  renormalization by \Ref{LevinNave} and \Ref{GW}.  \se{results} outlines the results from these calculations, including some new results for both the 1975 method and also the rewiring calculations. The final section suggests further work.

\section{The renormalization process}
\subsection{Overview}\la{overview}

We now come to compare different approximate real space renormalization schemes. The starting point for the considered methods is a system described by the statistical variables, $\{\sigma\}$, and a Hamiltonian $\mathcal{H}\{\sigma\}$. In the 1975 scheme this Hamiltonian is directly used to define the partition function
\bsubs \la{ZfromT}
\be Z=\text{Tr}_{\{\sigma\}}e^{-\beta \mathcal{H}}.
\ee
In the newer scheme,  the Hamiltonian  is used to define to define a two-, three-, or four- index tensor along the lines described in \se{Statistical} below.  The partition function is then defined as a statistical sum in the form of a sum over indices of a product of such tensors, in the form
\be
\text{~two index:~}Z= \text{Tr}_{i,j,k,...,n } ~~  T_{ij}T_{jk}.....T_{ni}
\ee
or
\be \la{four}
\text{~four index:~}Z= \text{Tr}_{i,j,k,...,  } ~~ \prod T_{ijkl}
\ee
\esubs
In both cases, the setup of the tensor product is such that each index appears exactly twice.
In this way, the system can maintain its gauge invariance  as an invariance under the rotation of each individual index.
We can then imagine that these partition functions may equally well be described in terms of the values of coupling constants, $\bf{ K}$, or of the value of tensors, $T$.

Working from this starting point, the renormalization scheme is implemented through three steps as follows:

\subsubsection{Introducing new statistical variables}
\begin{figure*}[t]
\begin{center}
\includegraphics[width=12cm ]{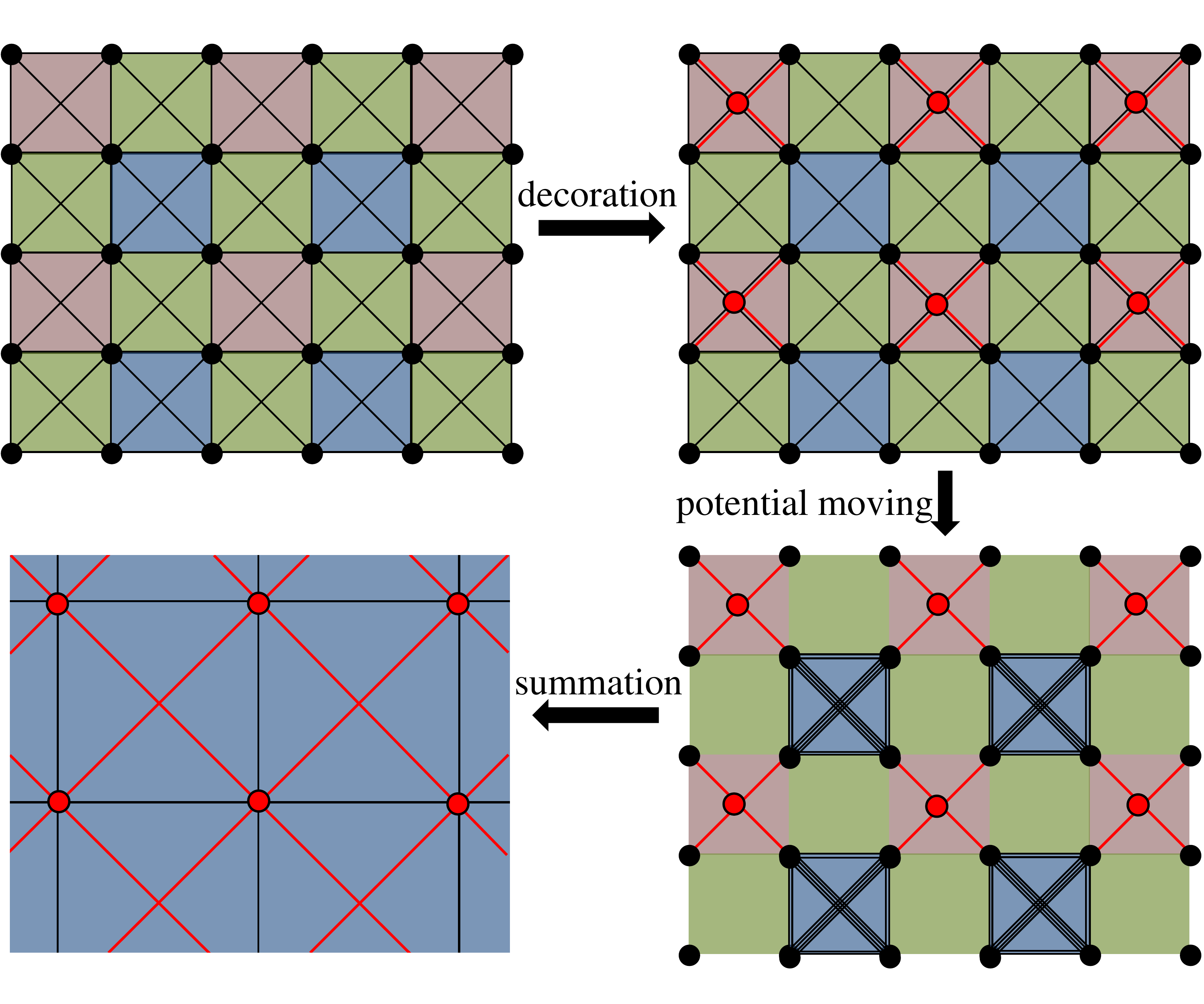}
\end{center}
	\caption{The setup for a potential moving scheme on a square lattice. The old spins ($\s$) are marked by filled black circles  located at the vertices of a square lattice. Note that each such spin belongs to four different squares. These squares form the ``blocks'' for our calculation. The new spins, $\mu$ appear in one quarter of the blocks, and are marked as filled bright red circles. The (red) lines emanating from these new spins denote coupling terms that link these to the old spins.  Each such coupling connects a single old spin to a new one. The potential moving places all the interactions between old spin in blue squares. The old spins around every blue square are connected only to themselves and to new spin variables. They can be summed over, giving a new effective coupling between adjacent new spins. }
\label{LB}
\end{figure*}

In the 1975-style scheme, the new variables are defined to be exactly similar to the old variables, $\{\s\}$, except that the new variables are spaced over larger distances than the old ones (See \fig{LB}.)\footnote{ In fact, this identity of old and new is one of the major limitations of the older scheme.}.  A new Hamiltonian depending on both old and new variables, is defined by adding to the old Hamiltonian  an
interaction term\footnote{The $\tilde{ }$ appears on this $V$ to distinguish it from another use of the symbol $V$,  that is the $V$ that conventionally appears in singular-value-decomposition analysis.}.
  $\tilde{V}(\{\mu\},\{\sigma\})$.
This term is defined so that the partition function remains unchanged by the inclusion of the $\mu$'s.  This invariance is enforced by the condition
\bsubs \la{Zconserved}
\be
\text{Tr}_{\{\mu\}}~~e^{-\beta \tilde{V}(\{\mu\},\{\sigma\})}=1
\ee
so that the partition function can be written as
\be
Z=\text{Tr}_{\{\s\}}e^{-\beta \mathcal{H}(\{\s)\}}=\text{Tr}_{\{\s\}}\text{Tr}_{\{\mu\}}e^{-\beta \mathcal{H}(\{\s)\}-\beta \tilde{V}(\{\mu\},\{\sigma\})}.
\ee
\esubs
A roughly similar analysis can be used in the tensor network scheme. Starting from the definition of the partition function as the trace of a product of tensors in \eq{four}   one replaces each of the rank four tensors by a product of rank three tensors, using a scheme derived from the singular value decomposition (SVD) theorem (See \se{subsec:svd} below.) as\footnote{The standard SVD scheme produces a matrix-multiplication product. $T=U \Sigma V^{Tr}$ where $\Sigma$ is a diagonal matrix. The diagonal entries in $\Sigma$ are non-negative and are called {\em singular values}.  In the notation in \eq{SVD'} the $\Sigma$ is absorbed into the $U$ and $V$.}
\bsubs \la{sumTensors}
\be \la{SVD'}
T_{ijkl}= \text{Tr}_\a ~ U_{ij\a} V_{kl\a}
\ee
 leaving us with
\be
Z= \text{Tr}_{ijkl...}~  \prod ~T_{mnpq}=\text{Tr}_{ijkl...}~\text{Tr}_{\a\b\gamma...} \prod ~ U_{ij\a} V_{kl\a}
\ee
\esubs

\subsubsection{Tracing out the original statistical variable}

In the 1975 scheme the trace over the original statistical variables, $\{\s\}$, defines a new Hamiltonian $\mathcal{H}'$ which depends solely on the new variables $\{\mu\}$,
\be
e^{-\beta \mathcal{H}'\{\mu\}}=\text{Tr}_{\{\sigma\}}e^{-\beta \tilde{\mathcal{H}}(\{\mu\},\{\sigma\})}
\ee
 so that
 \be
Z=\text{Tr}_{\{\mu\}} e^{-\beta \mathcal{H}'\{\mu\}}
\ee
One can expect that some approximation will be needed in order to calculate the sum over the $\s$'s.

A roughly analogous procedure can be applied to the tensor sums in \eq{sumTensors}.  If the position of the $UV$ products and the new indices have been deftly chosen, the old indices will appear in a series of small islands in which each island is only coupled to a limited number of new indices. Following \Ref{GW}, we shall work with the case in which that number is four. After a rearrangement, the partition function  sum in \eq{sumTensors} may be written as
$$
Z= \text{Tr}_{\a\b\gamma...} \text{Tr}_{ijklmn...} ~ \prod ~U(ij,\a) V(\a,kl)  ~  (\prod ~T_{mnpq})
$$
The sum over the old tensor indices may then be performed, generating new tensors, $T'_{\a\b\gamma\delta}$, so that
\be \la{new}
Z= \text{Tr}_{\a\b\gamma...}  ~  (\prod ~T'_{\a\b\gamma\delta})
\ee

\subsubsection{Obtaining a recursion relation}
The new degrees of freedom, $\mu$, have been defined to be identical to the variables, $\s$, the only difference being that the $\mu$'s are defined on a rescaled system. This identity usually permits the extraction of new coupling constants, ${\bf K' }$ from the new Hamiltonian. The new couplings are then connected to the old via the recursion relation $ {\bf K' } = {\bf \mathcal{R}(K) }$.

If the recursion relation is calculated exactly, the new set of couplings will  likely contain many more terms than the old set. This proliferation of couplings reflects the additional information from several blocks of the old system that we are trying to cram into one block of the new one.  An approximation is needed to limit the new couplings.  This limitation usually results in a situation in which  the possible couplings include only those that can be formed from spins completely within a geometrically defined block. Couplings which include spins from several blocks are excluded.   One example of such a block is shown in \fig{LB}.

The tensor scheme has a different approach.  In order to do renormalizations, the new partition function calculation of \eq{new} must have the same structure is the old one in \eq{four}.  As we discuss in detail in \se{subsec:svd} below, this structural identity is violated by the exact theory in which there are many more new indices than old. To obtain a recursion relation, one must use an approximation to eliminate the proliferation in the summation degree, $\chi$. As we shall discuss in \se{subsec:svd} below, an approximation of this kind is automatically provided by the SVD method. Using this approximation method, one has a renormalized problem with exactly the same structure as the original problem.  The result may be expressed as a recursion relation for the rank four tensor
\be \la{tensor}
T' =\mathcal{S}(T)
\ee
or as a recursion relation for the parameters defining those $T$'s, e.g. $ {\bf K' } = {\bf \mathcal{R}(K) }$.

There is a difficulty in using tensor components in the recursion relation of \eq{tensor}.  Because of the gauge invariance the components  of  the new tensor, $T'$ are not uniquely defined.  To ensure uniqueness, it might well be better to define the tensors in terms of gauge invariant parameters.  While this may be done relatively easily for low $\chi$ values, identifying all the independent gauge invariants for high $\chi$ value tensors may be a daunting task.

With a recursion relation at hand one may apply all the tools described in the previous section and obtain a fixed point Hamiltonian and the corresponding critical exponents.

In the remainder of this chapter, we describe the nuts and bolts of the real space renormalization process, using as our example square lattice  calculations based on Ising-models and the version of tensor renormalization found in \cite{GW}.  We particularly focus on understanding the differences between the older \cite{K75} and the newer styles \cite{LevinNave,GW} of doing renormalization work.

\subsection{Basic Statistical Description}
\begin{figure}[h]
\begin{center}
\includegraphics[height=3cm ]{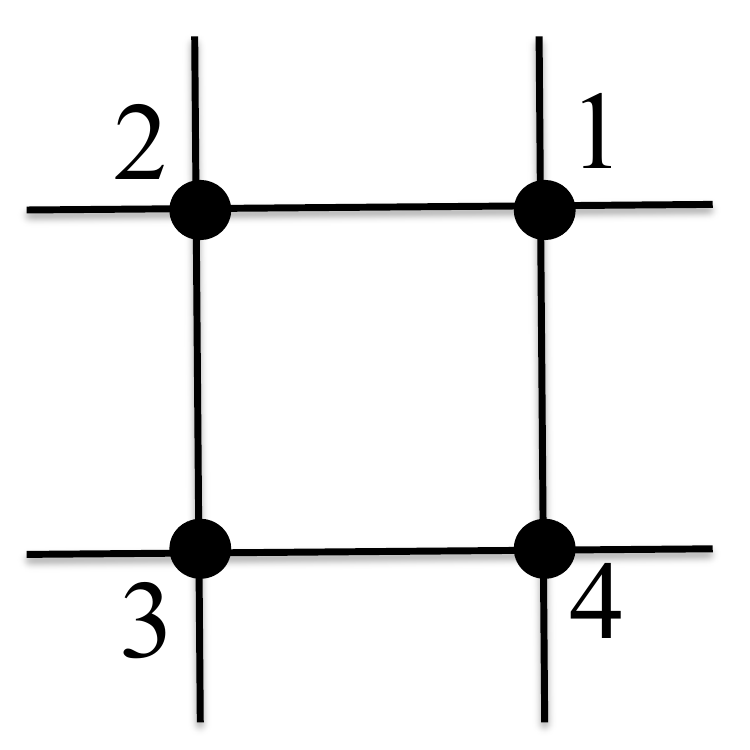}
\end{center}
	\caption{ Identification of the spin variables located at the vertices of a square unit cell.}
\label{fig:numberedSPIN}
\end{figure}

\la{Statistical}
In \se{Variables} we pointed out that the older calculations are based upon summations over defined stochastic variables like the Ising models $\s_\r=\pm1$. These calculations then use a Hamiltonian $\h(\{\s\})$ to define the statistical weight of each configuration of the variables.   Consider a problem involving four spin variables, $\s_1, \s_2, \s_3, \s_4$,  sitting at the corners of a square (see \fig{fig:numberedSPIN}), each variable taking on the values $\pm 1$. If this problem has the symmetry of a square, it can be described in terms of the following combinations

\begin{align}
S_0=&\, 1, \nonumber \\
S_1=&\, \s_1+\s_2+\s_3+\s_4,\nonumber \\
S_{nn}=&\, \s_1 \s_2+\s_2\s_3+\s_3\s_4+\s_4\s_1,\nonumber \\
S_{nnn}=&\, \s_1\s_3+\s_2\s_4,\nonumber \\
S_3=&\, \s_1 \s_2\s_3+\s_2\s_3\s_4+\s_3\s_4\s_1+\s_4\s_1\s_2,\nonumber \\
S_4=&\, \s_1 \s_2\s_3\s_4.
\label{eq:spin-variables}
\end{align}

The spin combination variables $S_i$ form a closed algebra, i.e. any function of the spin variables of \eq{eq:spin-variables}   may be expressed as a linear sum of these same variables with constant coefficients:
\[
F(S_0,S_1,S_{nn},S_{nnn},S_3,S_4)=\mathop{\sum} a_i S_i\,\,.
\]
One important example of this set of variables, denoted as $[S]$ is a  Hamiltonian $\h^{sq}[S]$ which describes the most general isotropic interactions with the symmetries of  a square unit block that can be formed from the set of $\s_i$'s .  The basic block used in the 1975 renormalization calculation is given in terms of this Hamiltonian as
\be
\text{BLOCK}= e^{-\b \h^{sq}[S]}  \text{~ with ~}  -\b \h^{sq}[S] = \sum_i  K_i S_i
\la{BLOCK} \ee
Here the $K$'s are called coupling constants and their values provide a numerical description of the problems at hand.

In contrast, a whole host of new calculations replace the coupling constants by tensors, and use the tensor indices as a proxy  for statistical variables. To illustrate this process, we write the tensor,  $T_{ijkl}$,  for the cases in which each index can take on two possible values and in which there is once more the symmetry of a square.  The tensors are situated on every other square and therefore capture only half of the possible four spin interaction and next nearest neighbor interaction.\footnote{ We note that allowing the index four possible values allows the description of every interaction in \eq{eq:spin-variables}.  However, in favor of simplicity we restrict our present treatment to the two valued index tensors only. }We use the spin notation to write the tensor as
\be
T =e^{\sum K_i S_i}.
\ee
There is considerable flexibility in defining the indices\footnote{In fact that flexibility is a sort of freedom under gauge transformations, and that freedom represents one of the main attractions of the tensor approach.}. For example, we could let one index-value, (+), correspond to positive spin and the other, (-), to negative spin. Then the tensor components would have the following distinct values

\begin{align}
T_{++++} = &\exp(K_0+4K_1+4K_{nn}+2K_{nnn}+4K_3 +K_4) ,\nonumber \\
T_{\{+++-\}} =&\exp(K_0+2K_1-2K_3 -K_4 ), \nonumber\\
T_{\{++--\}} = &\exp(K_0 -2K_{nnn} +K_4),  \nonumber \\
T_{\{+-+-\}} =&\exp(K_0-4K_{nn}+2K_{nnn} +K_4 ), \nonumber\\
T_{\{+---\}} = &\exp(K_0 -2K_1+2K_3 -K_4),  \nonumber \\
T_{----} = &\exp(K_0-4K_1+4K_{nn}+2K_{nnn}-4K_3 +K_4),
\label{eq:K2tensor}
\end{align}
where curly brackets stand for all cyclic index transformation, i.e.
\[
T_{\{+++-\}}=T_{\{++-+\}}=T_{\{+-++\}}=T_{\{-+++\}}.
\]
Alternatively,  one might use the index values $i =[1]$ to represent a sum over the statistical weights produced by the  possible spin configuration ($\s=+1$) and ($\s=-1$) and the index [2] to represent a  difference between these two statistical weights, specifically
\be
[1]= \frac{(+) + (-)}{\sqrt{2} }  \text{~and~}     [2]= \frac{(+) - (-)}{\sqrt{2} }  \la{transform}
\ee
The factor of $\sqrt{2}$ is introduced to make the index-change into  an orthogonal transformation.
Under this definition the tensor-representation would also have six distinct components however their values in the different representations change according to
\be
\tilde{T}_{ijkl}= O_{im}O_{jn}O_{ko}O_{lp}T_{nmop},
\la{IndexChange}
\ee
where $O$ denotes the orthogonal transformation which maps the indices $+-$ on the right  to the new indices $[1]$ and $[2]$ that appear on the left. For example:
\begin{align}
\la{indices}
\tilde{T}_{1111}=&\tfrac{1}{4}\bigl( T_{++++}+4 T_{+++-}+4T_{++--}\nonumber \\
&+2T_{+-+-}+4T_{+----}+T_{----}\bigr) ,\nonumber \\
\tilde{T}_{\{1112\}}=& \tfrac{1}{4}\bigl( T_{++++}+2 T_{+++-}- 2T_{+----}-T_{----}\bigr) .\nonumber
\end{align}

It is important to note that \eq{IndexChange} gives two different descriptions of the very same tensor, $T$, in different bases systems.  The tensors remain the same, but the coordinate system is varied.

Of course, the case described here is rather simple. The renormalization transformation develops, at each step, a succession of tensors, usually of increasing complexity,  At each step, the partition function depends upon the tensor in question, but is independent of the particular representation of that tensor.   When applied successively to the redefinition of indices in each step of a long calculation, the index method provides a flexibility and power not easily available through the direct manipulation of spin-like variables.  We shall see this flexibility in the specific calculations of renormalizations to be described in \se{results} of this paper.

\subsection{Tensor-SVD Renormalization} \la{Tensor-SVD}
In this section, we complete the discussion of renormalization as it was set up by \Ref{LevinNave} and \Ref{GW} and then carried out by  \Ref{Aoki} and \Ref{HB}. We begin with introducing the main tool of the method, the singular value decomposition, and discuss its properties. We then discuss the underlying geometry of the tensor network and review the tensor gauge freedom.

\subsubsection {Singular Value Decomposition}
\label{subsec:svd}
The new renormalization methods described in this paper are based upon the papers of \Ref{LevinNave} and \Ref{GW}.  (See also, for example \cite{Zhao,Xie}). These make use of  the singular value decomposition theorem in their analysis. The theorem states that every real matrix $M_{ij}$ can be expressed as a product of a real unitary matrix, a diagonal non-negative matrix and another real unitary matrix:
\be
M_{ij}=\sum_\a \Psi_{\a i} \Lambda_\a \Phi_{\a j},
\la{svd1}
\ee
where $\sum_\a \Psi_{\a i}  \Psi_{\a j}=\d_{ij}$ and $\sum_\a \Phi_{\a i}  \Phi_{\a j}=\d_{ij} $. While the decomposition is not unique the non-negative real numbers,  $\Lambda_\a$ are unique. Customarily, the  $\Lambda$'s, called {\em  singular values}, appear in descending order. When so ordered \eq{svd1} with only the $\chi$ largest components of  $\Lambda$ taken into account (the remaining set to zero), yields a rank $\chi$ approximation of $M$ which is optimal in the least square sense.\footnote{More precisely, the SVD estimate of $M$, called $M{^\chi}$,  serves serves to minimize the quantity $Q=\text{trace} (M-N)^2$ within the class   $N$'s that are matrices with only $\chi$ non-zero eigenvalues.  The minimizer is given by  $N= M{^\chi}$.} When $\chi=n$, the SVD approximation of \eq{svd1} is exact.

 For the specific case of a square $n$ by $n$ real matrix we may identify
\[
\Phi_{\a j}= \phi_{\a j},\quad
\Lambda_{\a}=|E_\a|,\quad
\text{and} \quad
\Psi_{\a j}= sign(E_\a)\psi_{\a j},
\]
where as above the $\phi$ and $\psi$ vectors denote the left and right eigenvectors of $M$ satisfying
\be
\la{EIGENtheorem}
M_{ij}= \sum_{\alpha=1}^n  \psi_{\a i}     E_\alpha   \phi_{\a j}.
\ee

 Finally, we may fold the singular value into the matrices $\Phi$ and $\Psi$:
\bsubs \la{SVDtheorem}
\be
 U_{\a i}  =  \Psi_{\a i} |E_\alpha|^{1/2}    \text{~   and ~}  V_{\a j } =  \Phi_{\a j} | E_\alpha |^{1/2}.
 \la{eq:PsiPhi}
\ee
The above matrices allow us to rewrite the rank $\chi$ approximation of $M$ as a product:
\be
M_{ij}\approx   {M{^\chi}_{ij}} =  \Sigma_{\alpha=1}^\chi  U_{\a i}    V_{\a j}
\la{eq:Prod-decomp}
\ee
\esubs
This approximation will be used throughout the discussion of rewiring methods.  Notice that the approximation in \eq{SVDtheorem} becomes exact when $\chi=n$. Also, the above decomposition of \eq{eq:Prod-decomp} is not unique.   As the columns of $\Psi$ are the normalized eigenvectors of $M M^T$, they have a sign ambiguity. One may lift this ambiguity, as we do in our following calculations, by setting the sign such that the first non-vanishing component of each eigenvector is positive. Note, however, that this resolution of the sign ambiguity is not invariant under a base change.

\subsubsection{SVD as an approximation method}
For a square lattice, one writes down the tensor product representation of the partition function as the trace over a product of rank four tensors
\[
Z= \text{Tr}_{ijklmn...}~  \prod ~T_{ijkl},
\]
in which each index occurs precisely twice.
The summation depends strongly on the topology of the network comprised of the indices connecting adjacent tensors. For this reason the usual methods of describing tensor calculations make heavy use of pictures.  We shall follow that precedent.

\begin{figure*}[t]
\begin{center}
\includegraphics[height=6.5cm ]{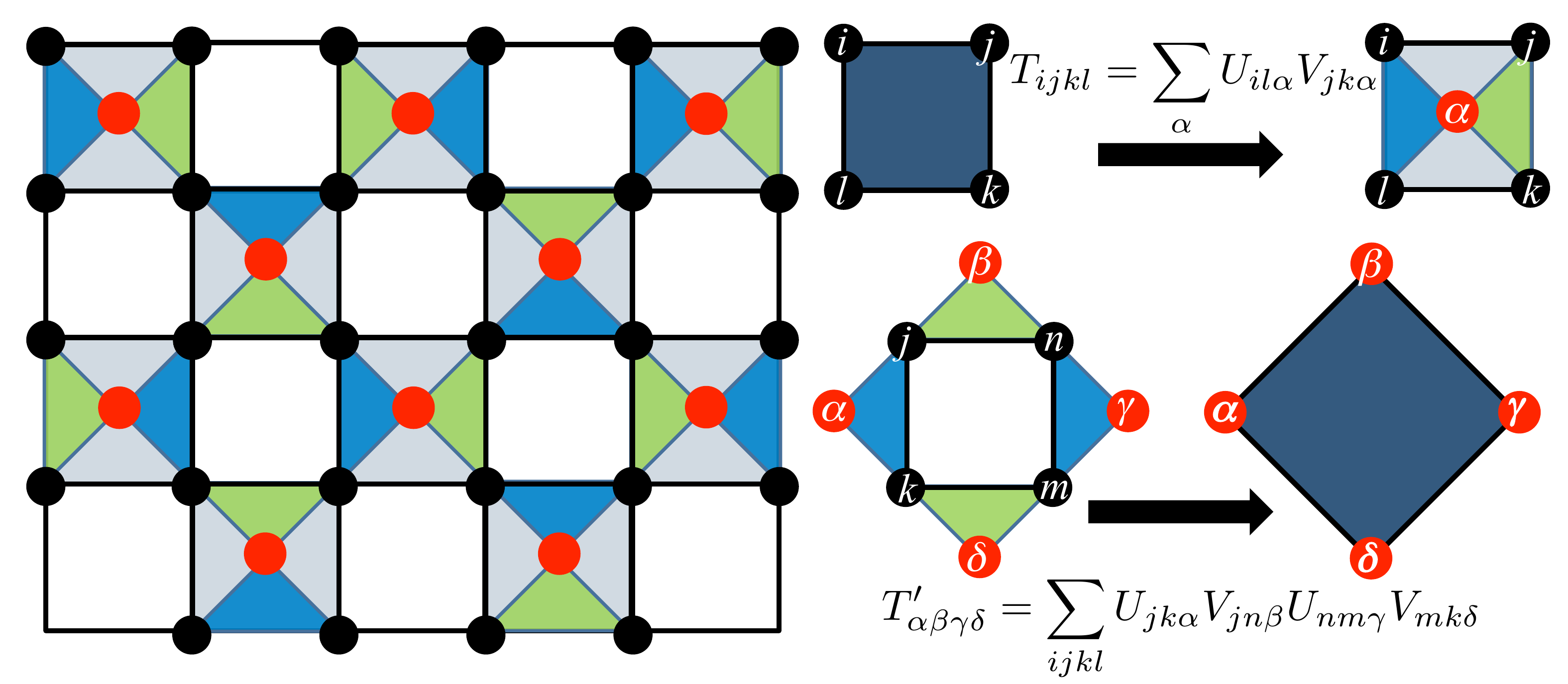}
\end{center}
\caption{The tensor network after rewiring.  The old four-legged tensors are shown lightly shaded.  They have disappeared and been replaced by the three-index tensors $U$ and $V$ respectively shown in blue and green.  Each three-index tensor appears as a triangle with two of the old tensor indices and one new index at its vertices. These are respectively shown as black and red  filled circles. Note the white squares.  These are all empty of interactions. These squares are of two kinds: the ones flanked by colored triangles (three legged tensors) and the one flanked by shaded triangles (the ghosts of disappeared four-tensors).  Each first-kind square permits the summation over the four old index variable at its corners and thereby the generation of interactions among the new indices. These white squares together with their four bounding triangles become the new tensors on the rescaled system. }
\label{ReWiredTensors}
\end{figure*}

 We show the tensor lattice in \fig{BasicTensor}.  Each colored box is a four-legged tensor.  The tensor indices appear at the corners as filled circles. The inset shows the definition of these indices.  The task at hand is to introduce new indices while isolating small groups of old indices so that these groups make no contact with other old indices.  To do this we rewrite a potential term like $T_{ijkl}$ as a matrix product in the form
\be
{T^{approx}}_{ijkl}=\sum_{\a=1}^{\chi'} U_{ij\a} V_{kl\a}
\la{SVD}
\ee
where $\a$ is the new index.  There are six ways  of doing this, involving different placements of the indices $ijkl$ in $U$ and $V$.  Two of these are depicted in \fig{ReWiredTensors}.

The singular value decomposition theorem points out that we can make \eq{SVD} give an exact expression for the four-legged tensor by using SVD and letting $\chi'=\chi^2$.
Alternatively we may use a smaller value of $\chi'$, as for example $\chi'=\chi$, and use either SVD or some other method to get a good approximation involving $U$ and $V$.  This kind of replacement is called a rewiring because it changes the connections in our lattice.  The change suggested by Gu and Wen\cite{GW} is shown in \fig{ReWiredTensors}. This figure shows that blocks of four old indices are coupled to new indices but not to any other old ones.  The four-index block draws its indices from four different tensors, T.   No interactions among indices are to be found in this kind of block before the renormalization process. All correlations are produced by the $U$'s and $V$'s that surround the block.  The calculation of the renormalized T is then very simple.  It is
\be  \la{SVDr}
T'_{\a\b\gamma\delta}=\text{Tr}_{jkmn} ~U_{jk\a}~V_{kn\b}~U_{nm\gamma}~V_{\delta mj}
\ee
In this way, a recursion relation is derived for any choice of $U$ and $V$.\footnote{ Note that the tensor $T'$ is not isotropic. Moreover, it is rotated by 90 degrees in adjacent cells. An alternative calculation resulting in rotationally invariant tensors sums either only $U$ matrices or only $V$ matrices for every $T'$. This results in two different tensors, $T'_1$ and $T'_2$, placed on a bi-partite lattice \cite{LevinNave}. }  However, it is natural and simple to use the SVD method to generate these three-legged tensors.  In the remainder of this paper we shall do that, fixing the number of indices by the condition $\chi'=\chi$.  In our numerical work, we shall stick with small values of $\chi$.   To get really accurate results, one squares $\chi$ several times until a large enough value is reached so that one feels one can neglect higher order indices.

\subsubsection{Gauge invariance and interpretation of fixed-point tensor components}
The tensorial formulation of a given statistical problem is, as previously mentioned,  not unique. In particular one may apply an orthogonal transformation to each of the legs of each of the tensors keeping the partition function obtained from their product invariant. One may naturally ask how does the fixed point tensor behave under such transformations.

We write the renormalization step \eq{SVDr} for a general tensor, $T$, as $T'=\mathcal{R}(T)$.
The fixed point tensor $T^*$ satisfies $T^*=\mathcal{R}(T^*)$. In the appendix we prove that up to sign ambiguities any rotation of the fixed point tensor yields under the renormalization step the original fixed point tensor, i.e.
\be
 \mathcal{R}(O_{im}O_{jn}O_{ko}O_{lp}T^*_{nmop}),=\mathcal{R}(T^*)=T^*.\footnote{This implies that a representation inependent definition of the fixed point tensor is given by  $\mathcal{R}(\tilde{T}^*)=\mathcal{R}(\mathcal{R}(\tilde{T}^*))$.}
\ee
This implies that there exists one particular ``preferred'' base for the representation of the fixed point tensor. The transformation that maps the summation variables from their physical representation to the fixed point ``preferred'' one is a priori unknown, and may be different for different fixed points.

While this makes the interpretation of the fixed point tensor non-trivial, in the majority of the SVD schemes such an interpretation is not necessary as fixed point analysis is not performed. Instead only the value of the partition function (and from it the value of free energy) is calculated as a function of a varying parameter, say the temperature. For a fixed point analysis, as we present here, one may wish to be able to cast meaning to the components of the fixed point tensor, and through them to the critical exponents.


\subsection{Null space of the response matrix}
The gauge symmetry discussed above implies the existence of a null space for the response function whose dimension is at least $\chi(\chi-1)/2$. Most of the works which analyze fixed point tensors employ gauge fixing
by treating a specific subset of the possible tensors.  In these theories the number of independent variables is greatly reduced, and they don't display the above null space.

Another contribution to the null space comes from the loss of information in the the truncation of the SVD decomposition. The number of independent components in a $\chi^2$ by $\chi^2$ matrix, in the general case scales as $d_1\propto \chi^4$. If however the $\chi^2$ by $\chi^2$ matrix is known to have only rank $\chi$ then the number of independent components reduces dramatically and scales as $d_2\propto \chi^3$. This implies a null space of dimension at least $d_1-d_2\propto \chi^4$. In the physical systems symmetries greatly reduce the number of independent tensor entries, and therefore also reduce the amount of information lost. 


\subsubsection{ Errors}
When $\chi'<\chi^2$ the approximate rewiring will generate an error.
We denote the local error resulting from the approximation by
\be
\text{Error}_{ijkl}=\ln[ T_{ijkl}/{T^{approx}}_{ijkl}]
\la{error }
\ee
This is the error of a single tensor at a specific configuration given by its indices values. The SVD scheme yields an error for the tensor that is optimized in a mean square sense \cite{LevinNave,GW}.  In the analysis of these authors, the error term is then simply neglected.
This method works exceptionally well for  large values of $\chi$, for which the error is quite small. In \se{results} and \se{discussion} we shall see that  this strategy does not work exceptionally well for smaller $\chi$.

An alternative approach to the neglect is to replace the error term by its maximum (minimum) over tensor indices. This yields an error of definite sign and in turn gives a lower (upper) bound on the free energy.   This approach can even be used to find  optimal values for $U$ and $V$ so as to give a best bound for the free energy.


The 1975 work employed a one parameter family of local lower bound approximations. The value of the parameter was carefully chosen such as to minimize the {\em global} error of the free energy, resulting in an error term that is quadratic in the local error term. In contrast the SVD scheme yields a free energy error that is linear in the error of \eq{error}.

This 1975 method proved to give plausible results for low $\chi$ values. We now turn to a discussion of this method.

\subsection{Lower bound variational renormalization}\la{lowerBound}
In this section, we complete the description of the lower-bound variational method. We first introduce the local conditions, formulated in terms of the symmetry of the Hamiltonian, which give rise to a lower bound on the free energy. We then construct a one dimensional family of such lower bound potentials characterized by a single parameter, $p$. We finally show how to choose the parameter, $p$, such as to globally minimize the resulting error in the free energy.

\subsubsection{Decoration}
As noted in section \ref{overview} the first step towards a renormalization is to introduce new statistical variables to the system, a process known as decoration.  In \eq{sumTensors} we described the tensor analysis scheme for doing the decoration.  Here we  describe  in more detail  the 1975 scheme for decoration.

In general each of the new degrees of freedom, which we will denote by $\mu$, is coupled only to a small subset of the old spin variables $\s$ through a coupling potential  $v([\s],\mu)$, (where $[\s]$ defines the small subset of the old spin variables). For example, in \fig{LB} every new degree of freedom is placed within an interaction block an interacts only within this block with its four nearest neighboring old spin variables. We define a new Hamiltonian within the interaction block of the new variable by
\be
\tilde{h}([\s],\mu)=h([\s])+v([\s],\mu).
\label{eq:HplusV}
\ee
where $h[\s]$ is the old Hamiltonian for the block.
Choosing the coupling potential to satisfy
\be
\text{Tr}_{\mu} e^{-\b v([\s],\mu)}=1,
\label{eq:Vcondition}
\ee
regardless of the specific value the variables $\s_i$, renders the partition function, and thus the free energy, unchanged by the inclusion of the new variable.   The full decoration is obtained by using
$$
\tilde{V}(\{\s\},\{\mu\})=\sum_\mathbf{R} v([\s]_\mathbf{R},\mu_\mathbf{R})
$$
where the sum over $\mathbf{R}$ is a sum over all $\mu$-sites. With the new Hamiltonian being $\tilde{\H}= \H + \tilde{V}$, the full partition function is unchanged by the decoration as in \eq{Zconserved}.

The 1975 scheme associates one new $\mu$-spin with the group of four old $\s$-spins in a surrounding square block.
There are multiple ways to choose a potential  interaction among the spins that will satisfy \eq{eq:Vcondition}.
Following  \cite{K75}, and \fig{LB} we define a one parameter family of such potential $v^p([\s],\mu)$, where the parameter $p$ serves to vary the strength of interaction amongst the new and old spins. This parameter will later allow us to optimize the choice of potential. The family of potential are given explicitly by
\bea
-\b v^p([\s],\mu)=&\, p \mu (\s_1+\s_1+\s_3+\s_4)+c([\s])\nonumber \\
=& p \mu S_1-\ln(2 \cosh(p\,S_1))
\eea
where $S_1$ is defined in \eqref{eq:spin-variables} and $c([\s])$ is chosen such that the sum of $e^{-\b v}$ over all values of $\mu$ gives unity.
Because of the closed form algebra of the isotropic spin variable \eqref{eq:spin-variables} we also know that the constant $c(\s)$ may be rewritten as a linear function of the isotropic invariant
\[
c([\s])=\sum a_i S_i.
\]
As a result, the  potential $v^p(\s,\mu)$ may be written as a linear combination of the $S_i$ values with coefficients which depend on the variational parameter, $p$.

\subsubsection{potential moving theorem}
The 1975 paper \cite{K75} employed a device for making the renormalization sum tractable that goes under the name of
{\em potential-moving}.  This device makes use of the following theorem:  Let us consider the statistical sum
$
e^{-\b F} = \text{Tr} e^{-\b \h}
$
where the trace gives a sum over a positive semi-definite set of terms involving a Hamiltonian $\h$, giving rise to a ``free
energy'', $F$. Now assume that $-\b \h=-\b \h_a+\delta V$.  Here, we shall use $\h_a$ to generate and approximate free energy, $F^a$
which we hope will have a value close to that of the exact free energy, $F$.  Our calculation makes use of the symmetry of
$-\b \h$ and $\delta V$,  in which we demand that $\delta V$ be odd under some exact symmetry of $-\b \h$, so that
\be
\text{Tr}[ e^{-\b \h} \delta V ]=0.
\la{symmetry}
\ee
This condition yields
\be
e^{-\b F^a} = \text{Tr~} e^{-\b \h_a}  \text{~~implies~~}  e^{-\b F^a} \geq e^{-\b F}
\la{partition}
\ee

To derive \eq{partition} define a Hamiltonian that interpolates between the exact and the approximate Hamiltonians and a free energy that arises from this interpolation.
$$
-\b \h(\lambda)=-\b \h_a+(1-\lambda) \delta V  \text{~and~}e^{-\b F(\lambda)}= \text{Tr} e^{-\b \h(\lambda)}
$$
These definitions imply that
$$ \frac{d}{d \lambda} ~ \b F(\lambda) =   < \delta V>_\lambda
$$
and
$$ \frac{d^2}{(d \lambda)^2} ~ \b F(\lambda) =  - [\delta V - < \delta V>_\lambda]^2_\lambda
$$
where the $\lambda$ subscript means that the average is calculated using a Hamiltonian $\h(\lambda)$.  It follows from \eq{symmetry} that the first derivative vanishes at $\lambda=0$.  The second derivative is always negative.  Therefore the interpolating free energy is always larger that the true free energy.  At $\lambda=1$ the interpolating free energy reduces to our approximate free energy.  Consequently,
\be
\b F^a-\b F = -\int_0^1~ d \lambda ~(1-\lambda)~ [\delta V - < \delta V>_\lambda]^2_\lambda \leq 0
\ee
Thus, the error in the approximation is of second order in $\delta V$ and that the approximate free energy provides an upper bound for the real free energy.

\subsubsection{Using potential moving}
To construct our approximate renormalization transformation, we need to make sure that the old spins are in isolated small groups, each group coupled to the new spins, but not to any other old spins.  If all the couplings obey this condition we can calculate the new approximate Hamiltonian.

We start from a situation in which the lattice is divided into square blocks as in \fig{LB}.  There are three kinds of blocks.
In figurative language, we think of $\delta V$ as containing some inconvenient couplings that interfere with our calculation of the
partition function in \eq{partition}.   What we do is then ``move'' the inconvenient couplings from their inconvenient
positions (in the new-spin blocks) to convenient positions in the other blocks. These convenient positions are required to be completely
equivalent in the exact version of the calculation to the inconvenient  sites so that \eq{symmetry} may be satisfied.  It is only our motion that produces the distinction between these two classes of sites.

The geometry of our calculation is shown in \fig{LB}. The original Ising spins appear as black dots at the vertices of
the squares. The new variables are the red dots in the red squares. The new spins are linked to the old spin variables by interactions indicated by the red bonds. Those bonds have interactions  of the form $e^{p\s \mu}$. All the squares have interactions described by
blocks of the form
\be
\text{BLOCK~}=\exp{(-\b \h^{sq})}
\ee
where the exponent is given by the block Hamiltonian using the stochastic variables defined by \eq{eq:spin-variables}.  In addition the red squares have a potential in the form of $c[\s]$ as given by \eq{BLOCK}.

All  the old interactions from the red squares and the green ones are ``moved'' into the blue squares. The potentials that exist at these squares define the motion. They are

\bea
 V &=& - \h^{sq}  \text{~ on green squares}  \nonumber \\
 V &=& - \h^{sq} +c([\s])  \text{~ on red squares} \nonumber  \\
 V &=& 3* \h^{sq} -c([\s])  \text{~ on blue squares}
\la{error}
\eea
The value of the potential on the blue squares is picked so that the sum of all the potential terms is zero, allowing for the double weight of the green squares.  The new interaction between the old and new spins can be formulated as a pairwise interaction between the new spin $\mu$ and each of its surrounding old spins. It therefore can be reformulated to be centered about the blues squares without any approximation. After the motion  of the potentials, we end up with no potential on green or red squares and a total potential   $4*\ \h^{sq} -c([\s]) $ on each blue square.

The error generated by the potential moving is proportional to the  mean variance in the $ V$ of \eq{error}.

After that motion the spins at the four  vertices of each blue square are linked to each other and to the surrounding new spins but to none of the other old spins.
This condition permits summations to be performed over each blue square independently of all the others, thereby producing interactions. The result is
\bea
e^{-\b \h'^{sq}([\mu])} =\text{Tr}_{[\s]} ~ \exp[&-4\b \h'^{sq}([\s]) -c([\s])\nonumber \\
&+p \sum_{j=1}^4(\s_j  \mu_j)]
\eea
The new coupling may then be projected out of the new Hamiltonian.  This projection then gives us the recursion relation.

\subsubsection{Spatial vectors and tensors}
This same mode of analysis enables us to discuss combinations of spin operators which behave like spatial vectors or tensors rather than the spatial scalers defined in \eq{eq:spin-variables}.  Thus, from the spin-labeling shown in \fig{fig:numberedSPIN}, it follows that the combination
$\s_1+ \s_4-\s_2-\s_3 $ is  to leading order the derivative of the spin with respect to the horizontal coordinate, ``x'', while
$\s_1* \s_4-\s_2*\s_3 $ is the derivative of the energy density with respect to x.   Similarly the two components $T_{xx}$ and $T_{xy}$ of the stress tensor\footnote{Of course, the stress tensor is a {\em spatial tensor} and not a gauge tensor like the $T$'s that appear in the rewiring scheme.} operator can respectively be identified as the simplest operators that have the right symmetry,
\bea
T_{xy} =& \s_1\s_3-\s_2 \s_4\, ,   \text{~and~} \nonumber \\
T_{xx} = &\s_3\s_4-\s_2 \s_3 +\s_1\s_2-\s_1 \s_4\, .
\eea
These identifications enable us to calculate the scaling properties of these operators in the lower bound scheme.  One simply calculates the scaling properties of these operator densities by putting these densities into the coupling of one particular blue square and then doing the recursion calculation for the lattice containing that one special square.  One can retain the lower bound property by setting up the potential-moving to be symmetrical about that square.        This approach then provides a recursion approximation for local operators that fit into a single block.  In the next chapter we show some eigenvalues for these vector and tensor operators.

No such scheme exists within the lowest order SVD analysis. Therefore we do not show eigenvalues for any vector or tensor operators within the SVD scheme.

\section{Results}\la{results}
Both the variational lower bound renormalization and the tensor renormalization can be realized by numerical schemes which produce fixed points and, more importantly, critical indices. The latter are expected to be a robust description of a critical point as they do not depend on the specific variables chosen to describe a given system. We next review some new numerical results, mostly in terms of critical indices, and compare them to others taken from the literature.

\subsection{Results from block spin calculations}
We review various systems differing in their underlying lattice structure (triangular, square and hexagonal), spin degrees of freedom ($\chi=2,3, \cdots$),   spin coupling (Ising, three state Potts and tricritical Ising) and methods of approximations. We collected the results in two subsections, separating the Ising models from the other models considered. Each subsection begins with a brief description of different systems and methods presented.
The critical indices of the various models are collected in two tables, concluding each of the subsections. Most of these results are not new.  They are results from 1975 \cite{K75,Bu,KHY}, somewhat augmented by calculations done for this paper. Additional results may be found in the literature (e.g. \cite{Bu5,Bu6,Nijs-Knops,Jan-Glazier}) , but the answers shown here are representative of the field.

The critical indices values, $x$, are derived from response matrix calculation at critical fixed points. The $x$'s are defined by
$$x=d-\log(E)/\log(\delta L)$$
 where $E$ is the eigenvalue of the response matrix at the fixed point, $d$ the dimensionality, and $\delta L$ the change in length scale. Each of the critical indices is associated with an eigenvector of the response matrix which represents a scaling operator, i.e. a linear combination of the system's operators which admits a simple scaling rule under renormalization \cite{Weg76a}.
The distinct indices were identified by the symmetry properties of their corresponding scaling operators and their values, as compared with exactly known results.
\subsubsection{Block spin renormalization of Ising models}
We list here different calculations using the block spin renormalization method as applied to the two-dimensional Ising model's critical fixed point.  The index-values are tabulated in \tab{MOVINGindices}, with the numbering given immediately below.
\begin{enumerate}
\item
The first-ever block spin calculation was performed by Neimeijer and  Van-Leeuwen \cite{N-van L}. In this work, studying Ising spins on a triangular lattice, the authors single out a subset of the triangular cells and separate the interactions into intracellular and intercellular interactions. The intracellular interactions are summed over and the intercellular interactions are recast as interactions between spins residing at the center of the chosen triangles. The ``unfavorable`` interactions which make the exact summation over intracellular variables intractable were simply neglected.
\item
The original lower bound model for Ising spins on a square lattice \cite{K75} as described above in section \ref{lowerBound}. The fixed point for which the critical exponents were computed exhibited equal nearest neighbor and next nearest neighbor couplings.
\item
The same system as above but at a different fixed point having unequal nearest neighbor and next nearest neighbor couplings.
\item The $\chi=3$ Ising model ( also known as the Blum-Capel model or spin 1 model) variational potential moving calculation of Burkhardt and Knops
\cite{Bu,Bu2,Bu3,Bu4}. This calculation generates once more a nearby pair of  fixed points: one with equal couplings between nearest neighbor and next nearest neighbor, the other with unequal couplings.  Their indices are sufficiently close to one another that they are not separately reported here.
\item  Ising spins on a hexagonal lattice studied via a variational potential moving calculation by Jan and Glazier \cite{Jan-Glazier}.
\item  Ising spins on a triangular lattice studied via a variational potential moving calculation by Jan and Glazier \cite{Jan-Glazier}.
\end{enumerate}

\begin{table*}
\begin{tabular}{l | c c c c c c | c }
lattice type&square& square&square&square&hexagonal&triangular& square\\
variational&no& yes&yes&yes&yes&yes&  \\
approximation-  &error-&potential-&potential-& potential-&potential-& potential-& none\\
method &neglect &moving &moving &moving&moving & moving& none \\
source &\cite{N-van L} &\cite{Southern}&\cite{K75,here}&\cite{Bu,here}&\cite{Jan-Glazier}&\cite{Jan-Glazier}&\cite{VCM}\\
\hline
  $x_{0}$~free energy& 0.0&0.0  &0.0 &0.0 &0.0&0.0 &0.0\\
  $x_{\s}~$ spin &0.12486 &0.12468 &0.12226 &0.1173&0.1289&-0.70.-0.31&0.1250 \\
  $x_{T}~$ energy &1.02774& 0.99912  &0.982473&1.0302 &1.0241&0.67,0.09&1.0  \\
  $\nabla \s$ & & 1.167 &1.073 &1.1440 &&&1.125\\
  $T_{xx}$ & & 1.797 &1.595&2.080&&&2.0\\
  $T_{xy} $ & &1.803  &1.595&1.569&&&2.0\\
  $\Phi$ &&1.79668  &2.11900  &1.98 & && \\
  $\nabla^2 $spin & & 2.06167 &  2.11689  &1.8303&&&2.125 \\
  $\nabla^2$energy & & 2.98391  &3.15848 &2.9389&&&3.0\\
\hline
\end{tabular}
\centering
\caption{Ising model critical indices. }\label{MOVINGindices}
 \end{table*}


In \tab{MOVINGindices} we list the critical indices obtained. The first three indices listed in the table are for the primary operators in the theory. Their values are known from the Onsager solution to the two-dimensional Ising model \cite{Onsager}, from the C.N. Yang \cite{Yang} calculation of that model's magnetization and from the results of conformal field theory \cite{FMS}[page 221].
The approximate numerical results for these indices are, with one exception, very close to the exact values.  The exception, the triangular lattice shown as number 6, displays indices that are considerably off the mark.  It has been argued \cite{Southern}  that the approximate calculation on a triangular lattice resembles a situation at a dimension different from two. However, that is an after-the-fact explanation.  We do not really know why the potential moving calculation does not work as well on the triangular lattice, or indeed why it does preform so well on the other lattices.

The next three critical indices correspond to higher order operators. These operator indices were not reported in the earlier papers and are first reported here. They are obtained by calculating the critical indices for operators that do not have the full symmetry of the BLOCK in  \fig{LB}. For example, the index of $\s_y$ is calculated from the recursion for the operator, $\s_1+\s_2 -\s_3 -\s_4$. Correspondingly, all of the operators beyond the primary ones are identified from their transformation properties under rotations, and can further be identified with the lowest order operators with the corresponding symmetry in the exact theory \cite{FMS}[page 221].

The operator marked $\Phi$ is a scalar operator which does not fit into the above pattern.  There is no operator with the corresponding index and symmetry in the exact theory \cite{VCM}[page 221].   Instead it is, we believe, a redundant operator \cite{Weg76a}, appearing essentially as an artifact of the particular method of normalization. A scalar operator with index close to 2.0 is a likely consequence of two nearby fixed points.  The existence of two fixed points is not required by the basic theory, and is itself a consequence of the particular method of constructing a renormalization.   Once one has two fixed points,  one expects to see an operator that powers the flow from one fixed point to the other.    As a redundant operator it is extraneous to the theory and has no ``correct''  index value.   But because it produces  changes in critical behavior, it is not surprising \cite{KW71} to see it has an index close to two.

The  $\chi=3$ result, depicted in the column numbered 4, disappointingly shows no better index-values than the ones in the $\chi=2$ columns. In fact one of the values, the one labeled $\nabla^2 \s $ is substantially worse. These results tend to suggest that one will not gain advantages from going to higher values of $\chi$ with the potential moving strategy as employed in the 1975 period. Perhaps this
result should have been expected.  The potentials moved  in the system will not become smaller for higher $\chi$ in this method.  This lack of convergence contrasts with what might be expected from rewiring calculations.   The rewiring calculations are believed to converge to the right results as $\chi \rightarrow \infty$, and that is their great virtue.

 \subsubsection{Block spin renormalization for other models}
Additional coupling constants appear when the spin variables are allowed to take more than two values. In this case, one can find, in addition to the Ising fixed point,  new fixed points displaying their own characteristic  critical behavior. We give in \tab{OtherIndices} a set of indices for the  fixed point
corresponding to the tricritical point of the Ising model and another set for the three-state Potts model. The critical indices values, calculated by potential moving methods on a square lattice,  are compared with exact values obtained from conformal field theory \cite[Chapter 7] {FMS}. As one can see, the agreement is not as good as the best obtained for the Ising model.  Nonetheless the values of the indices are good enough to be informative.

Here again a combination of the values of the critical indices and the symmetry of the corresponding operators were used to determine their identity. The free energy exponent is exactly zero in both the exact result and the approximate models.  The ``spin'' exponents describe operators that have the symmetry of the basic spins in the model.  Finally the operators marked energy display the symmetry of the fixed point.
Notice that the approximate calculations do not include all the indices available in the theory.  There are two sources of this omission.  The first is conceptual: If one starts with a limited set of operators, working with them will not necessarily produce all the operators in the theory.  This limitation particularly applies to the three-state Potts model.  The other limitation is  calculational.  The approximate calculations only produce meaningful results for a limited set of operators, those with the smallest values of the indices.

\begin{table*}[t]
\centering
\begin{quote}
\centering
\rule{-0.2in}{2in}\begin{tabular}{ l  c c c || c c  r }
 Tricritical Ising & Variational \cite{Bu,Bu5}& Variational \cite{Bu5}& \cite[page 222] {FMS}&Variational \cite{CD}& \cite[page 226]{FMS}&
3 state Potts \\
scaling operators& Potential moving & Potential moving &  Exact solution & Potential moving & Exact solution &scaling operators\\
\hline
 free energy& 0.0&0.0  &0.0 &0.0   &0.0&free energy \\
   spin 1 &0.0224 &0.0412 &0.0375&0.0896&0.0666&spin $\s$\\
    spin 2 & & &0.4375& &0.6667&spin Z\\
  \text{energy} $\epsilon$  &0.2030&0.1057&0.1  &&0.4 &energy $\epsilon$   \\
  \text{energy} $\epsilon'$  &0.8077 &0.8077&0.6 &1.1940&1.4 &energy X   \\
  \text{energy} $\epsilon''$  && &1.5 &&3 &energy Y  \\
  \hline
\hline
\end{tabular}
\end{quote}
  \caption{Critical indices for the three states Potts model and the Ising model tricritical point. The numerical calculations were preformed with $\chi=3$ on a square lattice and employed variational potential moving. }  \label{OtherIndices}
\end{table*}

\subsection{SVD Results}
We next review some fixed point results obtained for rewiring/SVD schemes at low values of $\chi$. The main analysis included is composed of critical exponents calculated at fixed points of the tensor renormalization scheme. Most tensor renormalization calculations find the critical indices from the free energy rather than  by calculating a fixed point.  Obtaining a fixed point is rather delicate in that it requires a careful treatment of the gauge invariance and also a careful control of which singular values will be included. These tasks, which are trivial for small $\chi$, become very difficult when successive iterations produce a large value of $\chi$.

We begin with interpretations of the configurations labeled by the different index values. We then summarize the results of the tensor renormalization calculation in  tables similar to  \tab{MOVINGindices} and \tab{OtherIndices}.

\subsubsection{Generation of SVD fixed point}
Our first SVD fixed point was generated for the $\chi=2$ square lattice Ising model. We put spins halfway along the bonds forming the legs of  the basic SVD tensor.  These spins are the red dots  in \fig{BasicTensor}.  For this square lattice each tensor has four legs.  Each such tensor can be described by a statistical weight
 $\exp[-\beta \mathcal{H}^{sq} ]$ .  We started from a tensor using two indices $(+)$ for up spins and $(-)$ for down spins.  We picked a tensor describing interaction strengths of the Onsager critical point of the two-dimensional Ising model.  Using the SVD recipe given in the end of subsection \ref{subsec:svd} we calculated an SVD decomposition  starting from a tensor defined  by
$$T_{++++}=T_{----}=e^{4K} $$
$$T_{\{+-+-\}}=e^{-4K} $$
with all the other tensor components having the value 1. (Once again, \{ . \} describes  any cyclic permutation of indices.)  This tensor represents the two-dimensional Ising model with nearest neighbor coupling $K$.  The statistical sum is a sum over products of such tensors, each having a weight determined by values of the four spins(See \eq{BLOCK}).
 We then performed SVD recursions at $\chi=2$, adjusting the tensor strengths until we reach a fixed point.

In the SVD decomposition \eq{svd1} the singular values represent the interaction strength between linear combinations of spin pair states.
The orthogonal transformations $\Phi$ and $\Psi$ in \eq{svd1} map the original spin pair states $(++),(+-),(-+)$ and $(--)$ to an alternative base which we denote by $[1],[2],[3]$ and $[4]$ with respect to which the spin-pairs interaction is diagonalized.\footnote{Whenever $\Phi$ and $\Psi$ are not equal then it is natural to decompose the lattice into a bipartite lattice of ``white'' and ``black'' sites. Each ``white'' spin pair interacts only with ``black'' spin pairs, and vise-versa. When diagonalized, every ``white'' spin pair displays a non-vanishing interaction with only one of the ``black'' spin pairs.} The rescaled tensor obtained after the renormalization step expresses the interaction between four ``new'' spins each of which attain one of two possible states; $[1]$ and $[2]$. This leads to a different representation than that we started from even if the new spin variable behaves exactly like an Ising spin.
For example,  half of the components of the fixed point tensor for  $\chi=2$  vanish (see \se{Tri2Ising}), whereas in the spin representation the tensor components cannot vanish. This is because the potentials involved are always finite and the tensor components in this representation are exponents of these potentials.

The components of the $\Phi$ matrix are given below in. It is helpful to note that both vectors, $[1]$ and $[2]$ are eigenvectors (with eigenvalues $1$ and $-1$) of the spin flip transformation, $+\to - $ and $-\to +$. We therefore define a linear combination of the two states with the desired spin flip symmetry:
\be
\tilde{[1]}=\frac{1}{\sqrt{2}}([1]+[2]),\qquad
\tilde{[2]}=\frac{1}{\sqrt{2}}([1]-[2]).
\label{eq:ortho}
\ee
It is easy to check that under a sign change of the original spin these new vectors transform as $\tilde{[1]}\to\tilde{[2]}$ and  $\tilde{[2]}\to\tilde{[1]}$. We identify these states as the ``$+$'' and ``$-$'' states of a new spin variable.

\begin{table}[h]
\begin{tabular}{ l | c c c c}
 \hline
  && SVD  index &&   \\
 &$[1]$ &$[2]$ &$[3]$&$[4]$ \\
  \hline
~~spin flip symmetry &even &odd&odd&even \\
~~interchange symmetry &even&even&odd&even \\
  \hline
  spin values\\
  \hline
$+ ~ +$ & 0.65708 &0.70711&0.0& 0.26124\\
$+ ~ - $&0.26124 &0.0 & -0.70711 &-0.65708 \\
$ - ~ +$ & 0.26124&0.0 &0.70711&-0.65708\\
$- ~ -$&   0.65708 &-0.70711 &0.0&0.26124\\
 \hline
singular value & 1.158 & 0.687 &0.109 &  0.064\\
 \hline
  \end{tabular}
 \caption{ The $\Phi$ matrix.  This four by four matrix describes the translation from a description that employs two  spin indices to one that employs a single tensor index.  This SVD translation is derived from the tensor for the  the $\chi=2$ fixed point of the SVD method. The different columns give the spin content of each index. For example, the index [1] describes a situation in which the $(+) (+) $and $ (- -)$ configurations have weight 0.66 while the other configurations have weight 0.26.The analysis is closed by an approximation that includes the first two index values, $[1]$ and $[2]$,  and neglects the other two. } \label{U}
 \end{table}

In the above scheme, a proper identification of the new spin variables after $N$ renormalization steps requires identification of each of the $N-1$ intermediary spin variables. In the example above, explicit calculation shows that the rotation described at \eq{eq:ortho} applied after every renormalization step retrieves the Ising spin variable. However, in general, this may not be the case and different rotations may be called for after multiple renormalization steps.
There is also an alternative approach. Instead of identifying the spin variable at every step in a path to a desired tensor, we examine only the desired tensor. We assume that it has some representation which could be interpreted as a spin representation and we solve for the orthogonal transformation $O$ mapping the given representation to the spin representation. In order to solve for $O$ we use the symmetries  of the tensor expected in the spin representation.

We carry out the renormalization steps without resorting to an identification of the different spin variables. We then find the fixed point tensor and seek a rotation $O$ such that in the rotated tensor we will be able to identify the physical couplings. It is important to state that as the rotation $O$ usually has limited degrees of freedom, finding a rotation which will make all the components of the fixed point tensor comply with the expected form may not always be possible. For example applying this logic to the $\chi=2$ square lattice fixed point described in \tab{U} and \se{Sq2Ising}, it is sufficient to require that $T_{++++}=T_{----}$  (expecting the fixed point to represent a zero field Ising critical point) to obtain that the 2D rotation $O$ in \eq{IndexChange} must be at an angle of $\pi/4 \pm \pi$ which is consistent with \eq{eq:ortho}. Below, in \se{Sq2Ising}, we use this identification of spin variables to interpret the response matrix in the vicinity of the fixed point.

   \begin{table}
\begin{tabular}{ l | c c c c c c}
 \hline
  &&& new index &&&  \\
 &$[1]$ &$[2]$ &$[3]$&$[4]$ &$[5]$ &$[6]$ \\
  \hline
SVDvalue & 1.22 & 0.81 &0.34 &  0.24 &0.07 &0.03\\
\hline
symmetries \\
~~spin flip &even &odd&odd&even&odd & even\\
~~interchange &even&even&odd&even &even  &odd \\
  \hline
  old indices\\
  \hline
$[1] ~ [1]$ & 0.86 &0        &0          &0.158 &0  &0\\
$[1] ~ [2] $&0     &0.69     & 0.43    &0 &0.16&0 \\
$[1] ~ [3]$ & 0    &0.16     &0.55     &0&-0.69&0\\
$[2] ~ [1]$&   0   &0.69     &-0.43       &0 &0.16&0\\
$[2] ~ [2]$ & 0.45 &0        &0           &- 0.56&0.16&0\\
$[2] ~ [3] $&-0.13&0          & 0           &-0.56 &0&0.71 \\
$[3] ~ [1]$ & 0&0.16        &-0.55        &0&-0.69&0\\
$[3] ~ [2]$&   -0.13 &0      &0             &-0.56&0&-0.71 \\
$[3] ~ [3]$&  - 0.16 &0      &0              &-0.20&0&0 \\
 \hline
  \end{tabular}
  \caption{The  $\Phi$ matrix for the $\chi=3$ fixed point. The transformation from two SVD indices to a single index.  The $\chi=3$ renormalization only
uses columns $[1]$ - $[3]$ of the table. The last three columns are not shown because they have relatively little influence on the $T$-matrix since their  singular values are 0.02, 0.003, and 0.002.}
 \label{UV3}
 \end{table}

\subsection{Summary of SVD numerical results}
In this section we list and describe some critical indices generated by  low order ($\chi=2,3,\text{ or }4$) SVD calculations.
\subsubsection{Hexagonal $\chi=2$}\la{Tri2Ising}
The simplest SVD calculation is on a hexagonal lattice.  The renormalization increases the lattice constant by a factor of $\sqrt 3$.  As in all the calculations described below, one finds the fixed point by starting out with a tensor representing a spin-flip-symmetric  triangle, invariant under rotations through $120^\circ$.  In the spin representation,  this situation is represented by the two couplings: $K_0$, a normalization constant, and the nearest neighbor coupling $K_{nn}$.  In the SVD representation generated from this one, there are two independent  tensor components, $T_{111}$ and  $T_{\{122\}}$.

There are two trivial fixed points:  A high temperature point in which $T_{111}$ and  $T_{\{122\}}$ both equal unity, and a low temperature fixed point in which $T_{111}=1$ and  $T_{\{122\}}=0.$

The critical fixed point is first found by searching in the ``space'' formed by the ratio of these tensor-components.  After many recursions most starting points will lead to one of the trivial fixed points.  However, between these two possibilities, one starting point with $T_{212}/T_{111}= 0.52454857$ will give a non-trivial fixed point. Two couplings means two  critical indices.  The exponents read zero for the free energy, and $x_T=0.98457$, for the temperature or energy.

To go further, one can include couplings describing configurations that are odd under spin flip.  As one can see from \tab{U} and \tab{UV} there are two groups of tensor elements of this kind,   $T_{\{112\}}$  and $T_{222}$.  These four tensor-components are set to zero at the fixed point. Including these components in the response analysis gives two more eigenvalues, $E=2.5549$ and $E=0$, which then generate the $x$-values 0.2923 and $\infty$.   These data are stored in the second column of \tab{SVDindices} along with data from our other SVD fixed points.

Because the $U$-values in \tab{U} and \tab{UV} indicate this this calculation has the symmetry of an Ising model one can immediately identify the fixed point just found as an approximate representation of the two-dimensional Ising model.  The index-values in part  support this identification.  The first scalar indices $x=0$ is exactly right.  The second, $x_T=0.98...$ is satisfyingly close to the exact value, 1.0. The spin index is, however, more than a factor of two larger than the exact value 0.125.  On the other hand, the infinite value of $x$ is exactly what we would expect from the gauge freedom built into the possibility of rotations between our two index values.

In calculating the recursion relation for the odd-in-spin-flip couplings, we found a difficulty that had to be surmounted. The second index, [2], could change its meaning as a result of very small perturbations. Its sign was essentially undefined.  Since the tensor components with an odd number of [2]'s are all zero at the fixed point, such a sign change might be considered to be ``no big deal''. However, a sign change engendered by an almost infinitesimal change in the tensor components defining the SVD transform can make a big difference in the calculation of the derivative of a recursion relation.  That in turn can ruin the calculation of a response matrix.  This kind of difficulty can be surmounted by defining the ambiguous signs in the $U$-matrix {\em ab initio.}

\subsubsection{Square $\chi=2$}\la{Sq2Ising}

The story is only a little different for the $\chi=2$ case on the square lattice.  The  fixed point tensor has the following non-zero fixed-point elements $T_{1111}=0.98669$, $ T_{\{1212\}}=0.28904$,
$T_{\{1122\}}=0.39757$ and $T_{2222}=0.2357$.  The indices have the same symmetry properties as the hexagonal case, so we can once more identify the situation as the critical point of the two-dimensional Ising model.   These even-under-spin-flip elements generate four eigenvalues:
$x=0$ for the free energy, $x_T=0.98331$ for the temperature,  x= 5.56 for some unidentifiable operator, and $x=\infty$ reflecting the lossy nature of the SVD truncation.

When one adds an analysis of the two different classes of odd-under-spin-flip tensors one gets two more eigenvalues, $x_\s= 0.25848$ and $x=\infty$. The last index is reflects a zero eigenvalue produced  by the freedom to rotate indices.  Except for the two large indices, which  presumably do not reflect any critical property of the Ising model, all this look surprisingly similar to the triangular case.  Once more, we have a pretty good  approximate representation of known Ising result, marred by an unexpectedly bad $x_\s$.

\subsubsection{Hexagonal $\chi=3$}\la{Tri3Potts}

One of the hexagonal lattice the $\chi=3$ fixed point tensors has a structure determined by two non-zero tensor components,
\be
T_{111} \text{ as well as } T_{\{122\}}=T_{\{133\}}
\la{Potts}
\ee
The structure of both of this tensor and of the $U$-matrix indicate a full symmetry between the configurations by  [2]  and by [3].
At the fixed point, the ratio of these tensors is  $T_{\{122\}}/T_{111}=T_{\{133\}}/T_{111}
=0.57735027$. The identical behavior of the [2] index and the [3] index is a reflection of the basic symmetry of this situation.
A further indication of this symmetry is the singular values, which are identical for for these two indices.  This behavior can be expected from the three-state Potts model, in which the system can line up in any one of the three components of its spin-variable.  There are then two linearly independent orderings.  This degeneracy is reflected in the possibility of rotations of the indices [2] and [3] into one another.

This degeneracy of singular values made the calculation of a fixed point and the evaluation of a response matrix and of eigenvalues very hard. The problem was solved in part by artificially breaking the [2]-[3] symmetry, for example by making tensor components containing the index [3] differ from ones with the index [2] by about one part in $10^8$, and then seeing what response eigenvalues might arise.  Two eigenvalues appeared robustly, ones with $x$-values of zero and 1.4322741.  The zero is, of course, the expected response of the free energy, while it seems reasonable to identify the latter values with the operator $X$ of the three-state Potts model\cite[page226]{FMS}.  That operator has an $x$-values of 1.4.

The analysis is, however, highly unstable and often shows an $x$-value of 1.00740.  This could very likely be a reflection of the thermal index of the Ising model.  That model is what should arise from the broken symmetry that we artificially added.

\subsubsection{Additional fixed point (hexagonal $\chi=3$)}\la{Tri3Potts2}

We found a second fixed point that, at first sight, seemed qualitatively similar to the Potts model fixed point described in \eq{Potts} above. It looks as if we are heading once more for a fixed point of the three state Potts model. However, in this case the one non-zero fixed point ratio is  $T_{133}/T_{111}=T_{122}/T_{111}=0.7689453$. Thus the coupling is much stronger than in the previous case.

Furthermore, the $x$-values are not at all the same as in the previous case.  In addition to the ubiquitous $x=0$, we find $x$=0.02669, 0.02751, and   2.94810.  The last $x$-value is likely to belong to the Potts operator called $Y$ that has the exact $x$-value of 3.0.  A possible identification of the previous two is with the Potts ordering operator with $x_\s$= 1/15 = 0.06667 in the exact theory. However, there are six operators in the theory \cite[Chapter 7] {FMS} that should all be generated in an algebra containing spin operators.  Thus the description we have given here is not very satisfactory.

\subsubsection{And one more (hexagonal $\chi=3$)}\la{Tri3Ising2}
The non-zero fixed point tensor values are
\bea
T_{111}=0.9955388,\nonumber \\
T_{\{133\}}=0.5210051,\nonumber \\
 T_{222}=1.0004890. \nonumber
\eea
This situation is a direct product of a critical point of a $\chi=2$ Ising model (indices [1] and [3]) and a trivial $\chi=1$ situation (index 2).  It then has two indices close to zero and also indices  0.3060 and  0.9868 reflecting the Ising model as well an additional uncoupled model in a trivial fixed point.

\subsubsection{Square $\chi=3$}\la{Sq3Ising}

On the square lattice, there is at least one  fixed point tensor for $\chi=3$ that describes the two-dimensional Ising model.  This tensor has ten different kinds of components, each with its own separate value.  However, the outcome of the response analysis is entirely familiar.  The three lowest $x$-values are zero, for the free energy, 0.27573 for the     magnetization , and
0.985346 for the thermal index.   We should notice that neither the higher  $\chi$, nor the additional complexity of a four-index tensor has yielded any improvement (or change) in the response eigenvalues.

\subsubsection{Square $\chi=4$}\la{Sq4Ising}
Aoki and coworkers \cite{Aoki} have calculated the $\chi=4$ fixed point on the square lattice, getting almost exactly the same values of the free-energy and thermal indices as they obtained for $\chi=3$.  They did not report a value for the magnetization index.  Once again we might feel disappointment to see that additional complexity did not produce improved accuracy.

\subsubsection{Four-state Potts model}  
We found, but did not analyze, several $\chi=4$ fixed points.  One of these is especially worth mentioning. This one describes a situation that appears to be trying to represent the four-state Potts model, but does not quite get there.  It has one configuration, [1] that represents a scalar background situation with a tensor component tensor $T_{111}=1.0093788$. In addition there are  three other indices [2],[3] and [4].  Our approximate numerical fixed point makes the tensor-components  described by these indices almost equal in value, viz

\bea
T_{\{122\}}=0.4492365 \nonumber \\
T_{\{133\}}=0.4492366 \nonumber  \\
T_{\{144|\}}=0.4492367 \nonumber
\eea

Finally, all six of the $T_{ijk}$'s that contain all three of the higher index values  (e.g. $T_{234} $)  have the value 0.3515106. Note that the values of three of the independent tensor entries differ from each other by less than one part in a million. This difference is, however, important in order to obtain a fixed point to numerical accuracy. Equating all three independent entries above does not result in a fixed point.
However, adding the same small constant   ($\sim 10^{-6}$)  to these three independent entries (while keeping them distinct) results in an equivalent fixed point, to numerical accuracy. This
 numerics reflects the null space of the response matrix at the fixed point.  It also points to the numerical delicacy of the calculation in the presence of multiple $x=0$ critical index values.

Note the even spacing of the numerical values of the magnitude of the $T$'s.  The reason that three index values represent four possible values of the Potts-model ``spin''   variable lies in the fact that, from the four probabilities of having one of four different values,  one can form three linearly independent difference variables.  In addition there is one trivial variable, the sum of these probabilities, that then has the value unity.

Unfortunately, because of the near-degeneracy of this situation, we have had difficulties analyzing the consequences of this model.  The near-degeneracies make the eigenvalue analysis, both in the response and in the SVD much more difficult to understand.

 \begin{table*}[t]
 \centering
\rule{-.750in}{0in}
\begin{tabular}{ l c c c c c c r }
lattice type&hexagonal&square&square & hexagonal&hexagonal&hexagonal&square\\
 $\chi$ & 2 & 2 &3 &3 &3 &3&4\\
source & \cite{here} & \cite{Aoki,here}  & \cite{here} &\cite{here} &\cite{here}&\cite{here}&\cite{Aoki,here}\\
 \hline
  fitted &Ising&Ising&Ising&Ising&q=3 &q=3&Ising\\
  model&&&&$\chi=1+2$&Potts &Potts\\
  \hline
  free energy& 0.0&0.0  &0.0 &0.0   &0.0&0.0 &0.0\\
 & & & &-0.0018&  \\
   $x_{\s}~$ spin &0.2924 &0.25848 & 0.27573 &0.30605&0.02752&-&0.32202\\
& & & & &-0.02669   \\
  $x_{T}~$ energy &0.9846&0.98330 &0.98534 &0.98685&-&1.43227&0.98330   \\
  $Y$ &&&&&2.94810 \\
  \hline
  \end{tabular}
  \caption{Primary Results from rewiring calculations using SVD.  The index values are derived from fixed point calculations that hold on to indices that have the same symmetry as the lattice.  The energy,spin, etc. are defined to be the scaling operators with the appropriate symmetry and the smallest x-value. As mentioned in the text, we have left out some of the larger response eigenvalues.  These are apparently not meaningful in the SVD calculations.}
 \label{SVDindices}
 \end{table*}

\subsubsection{The response matrix in tensor renormalization}
A first step In doing a calculation of the response matrix for the SVD scheme is to ask how many finite (non-zero) eigenvalues one might expect.  This counting is not a simple task.  For example, in the square lattice the renormalization is based upon a fourth order tensor, and that tensor has $\chi^4 $ different components.  For $\chi =2$, these sixteen component can be reduced to six by demanding that the tensor have the symmetries of a square.  Of these six, four ( $T_{1111},\,T_{\{1122\}},\,T_{\{1212\}},$ and $T_{2222}$) are even under spin flip, and two ($T_{\{1112\}}$ and $T_{\{1222\}}$) are odd.  However, gauge symmetry allows rotation of  the two indices, eliminating one of the odd non-zero tensor components. So we expect four finite even-spin $x$-values, on such odd spin value, and one infinite $x$-value corresponding to the gauge symmetry.

\begin{figure}[h]
\begin{center}
\includegraphics[height=6cm ]{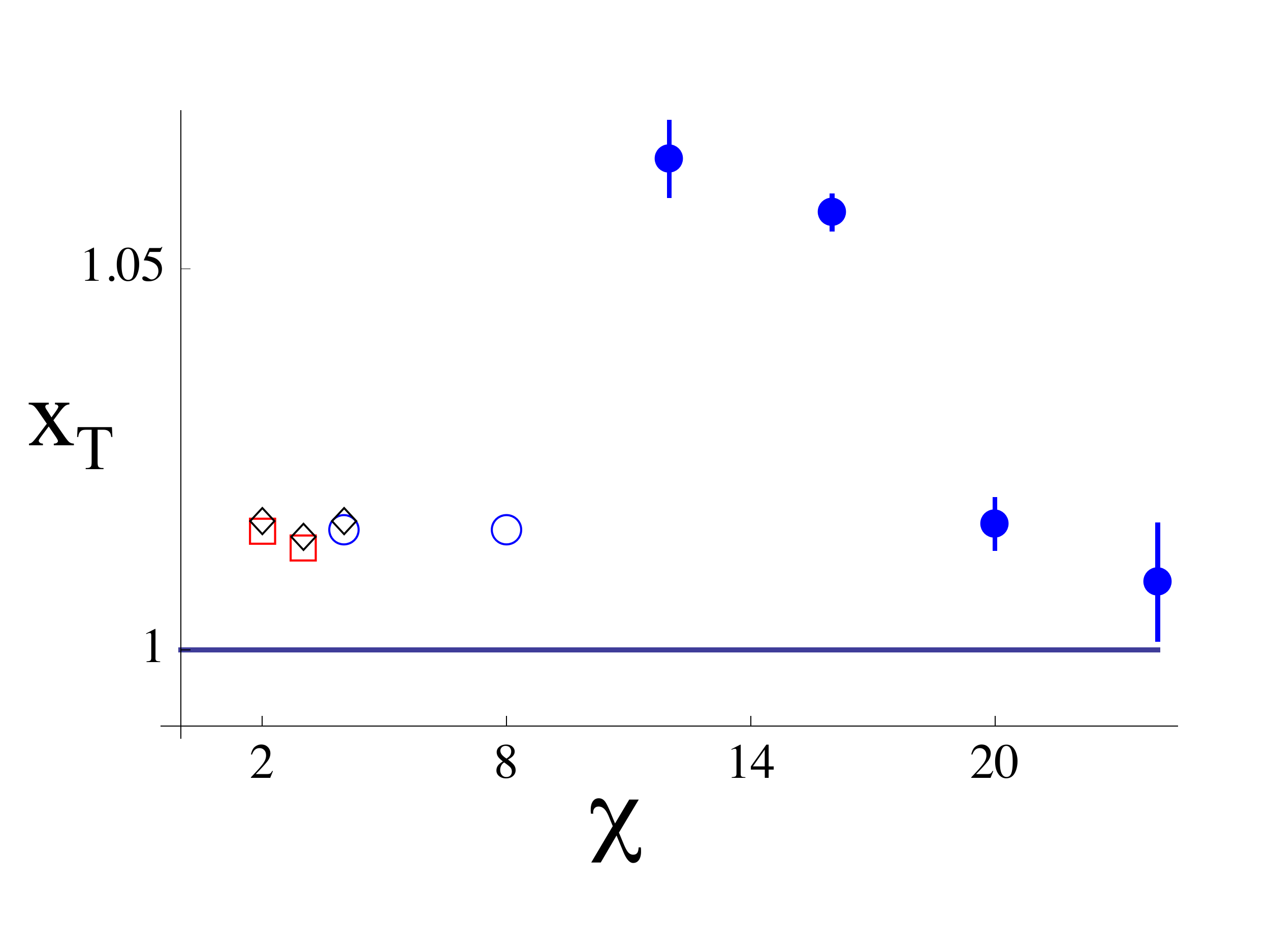}
\end{center}
\caption{Thermal x-value versus $\chi$. (Blue) Circles denote the results described in reference \cite{HB} obtained for an hexagonal lattice tensor product. (Red) squares denote the exponents for a square lattice obtained here, and Diamonds denote the results for a hexagonal lattice obtained here. All filled circles denote exponents obtained through the fitting of the free energy in the vicinity of the critical point. The empty markers denote results obtained directly from the calculation of the response function of the recursion relation around the fixed point. }
\la{xT}
\end{figure}

\section{Discussion} \la{discussion}

\subsection{Error estimates}

The error in a square lattice rewiring  calculation is proportional to the deviation from unity of the ratio of the exact four-index tensor to the
approximate one used in the analysis. A wide variety of methods can be used to obtain the approximate tensor. We follow previous workers and employ SVD analysis here.  In any step of the SVD analysis, the error may be set to zero by choosing the new value of $\chi$ to be the square of the old value.   However, the computational complexity of the calculation will, at some point, have to be limited by demanding that the increase in $\chi$ will cease.   If this eventual value of $\chi$ is large, one can expect the calculational error to be small.   Since the  error terms are simply neglected in the analysis, we
might expect that the  inaccuracy in critical indices should be linear in the error.  For this reason, we should not be surprised
if low values of $\chi$ gave inaccurate results for the recursion within the SVD analysis, but that the calculated  free energy value should converge quite
satisfactorily for larger values of $\chi$ \cite{VCM}.

On the other hand, our numerical results for critical indices might suggest a different story.  The magnetic index taken from the fixed point for the Ising model and its cousins is uniformly in error by about a factor of two.  In contrast,  the thermal index starts out, for small $\chi$,  accurate to within a few percent and then seems to  slowly improve its value for higher $\chi$. (See \fig{xT}.) Might we have a situation in which the free energy derived from the fixed point converges quite well, but the indices do not show equal convergence.   There are some hints in \cite{LevinNave} and  \cite{GW} that they expect much better convergence of the free energy away from the critical point than at that point.  The whole effort to use SVD for the calculation of fixed points might be fraught with conceptual difficulties.

The SVD-rewiring fixed point calculations show a wide variety of numerical difficulties.  The most natural way of finding a fixed point involves a Newton's-method search.  That approach, in turn, requires that the approximation used give the parameters that determine the fixed point in differentiable form.  However,   there are several important impediments to such differentiability, including
\begin{itemize}
\item Crossing of singular values.  The SVD method does not necessarily make the approximate matrix be analytic in the parameters of the approximated one. In particular, singular values may cross one another, producing a result containing discontinuous derivatives.    This problem is likely to result in very delicate numerics for large values of $\chi$.
\item Degeneracy of singular values.  In situations with higher symmetry than the Ising model, the singular values may be degenerate, making the SVD calculation very sensitive to small perturbations.
\item Gauge symmetry.  The approximate matrix will have a gauge symmetry that makes some combination of components of the approximate tensor insensitive to components of the tensor being approximated.  This effect then produces an indeterminacy in the output, and thus a pathological sensitivity to numerical errors.
\item Order parameter.   The order parameter does not fit smoothly into the SVD scheme. The critical system will fluctuate among several states of order.  In our calculations, and probably in all SVD calculations, some index variables had $U$ and $V$ values that varied discontinuously as one went from one state of order to the other.  As a result we saw discontinuous derivatives of the recursion matrix.
\end{itemize}
Whatever the cause, the net result is that, for statistical mechanical problems as distinct from Hamiltonian ones,  as far as we know, nobody has calculated fixed points for $\chi$ beyond 8.   Further, the one reference that has gone to large $\chi$,  \cite{HB}, sees an $x_T$ that shows  little improvement as $\chi$ increases in this range\footnote{For $\chi=2,3,4$, we find that this critical index is $0.985\pm 0.0015$ compared with the exact value, 1.0.  For higher $\chi$, estimates in \Ref{HB} give the disappointing value $x_T=0.938 \pm 0.005$ at $\chi=12$,  and the more pleasing value $0.991\pm 0.007$  at $\chi$=24.  Convergence is  slow and erratic.}.  (See \fig{xT}.)

In contrast, the potential-moving scheme, factored into a renormalization calculation,   has given remarkably accurate
results for simple models of critical behavior \cite{K75,Bu,Bu2,Bu3,Bu4,Knops,Nijs-Knops,KHY,Jan-Glazier,Dasgupta,Katz}. (See \tab{MOVINGindices} and \tab{OtherIndices} which lists critical indices for the two-dimensional Ising model as derived from this kind of  analysis.  Both thermal and magnetic critical indices derived in this manner are remarkably accurate. Several workers ( for example see \cite{Bu3,Southern,NIJS} ) expressed surprise about this high accuracy.  Additional indices, also
listed in the table are qualitatively reasonable.

There are several reasons for the increased accuracy of potential-moving relative to the SVD scheme with a similar (small) value of $\chi$.  Once again the source of error may be measured as a four-index tensor, here the tensor that defines the various potentials to be moved.   However, in this case, because the first order effect of the motion vanishes at $\lambda=0$, the inaccuracy in the free energy must automatically be second order in the error-source.  This change is the first reason for the improvement over SVD.

In addition, the parameter is adjusted to produce a minimum
change in free energy.  This adjustment pushes the error-source to be as small as it can be.  Further, the error source is
required to have a lattice-average that is zero.   This means that it does not behave as a spatial scalar but rather as a spatial tensor.  In the
calculations described here,  the symmetry of this perturbation is the same as that of the tensor $T_{xy}$.  For this reason,
all scalar quantities will be unmodified in first order, and will only see the direct effects of the potential-motion at second
order.

Nonetheless, our use of the potential-moving scheme has serious flaws.  The most serious one is that we do not know how accurate the method might be.  Sometimes it behaves better than expected, sometimes worse.  In addition, we know nothing about convergence at higher values of $\chi$.

\subsection{Work to be done.}
One can hope that the methods of analyzing the rewiring can be improved.    We would argue that the advances to be considered might include
\begin{itemize}
\item {\bf  Avoid gauge degeneracy.}  One should calculate renormalizations and recursions using gauge-invariant quantities, built for example from traces of the tensors. This will eliminate the worst source of numerical instability.
\item {\bf  Understand gauge degeneracy.}  The authors of this paper do not understand the reason that gauge degeneracy should underlie these statistical mechanical calculations.   A deeper understanding might bring us to better control of the method.
\item {\bf Control index degeneracy.} A physical symmetry can give a degeneracy in SVD and response functions.  Learning to deal with these can be a great help.
\item {\bf The magnetization.}  We don't understand why this index is much less accurately determined than the thermal index. We should understand that.  If we can, we should design an alternative way of determining this index.
\item {\bf More global analysis: Construct a variational scheme.}  We need a calculational scheme to replace SVD.  One possibility is to replace $U$ and $V$ by arbitrary three-legged tensors and minimize the free energy error they produce. \footnote{We note that alternatives to the choice of largest $\chi$ singular values, which are optimized globally rather than locally, as proposed in \cite{Zhao}, when restricted to low $\chi$ values did not result in improved exponents.}  One may achieve this goal following the principles of the variational potential moving. However, the real challenge is to design a scheme that both improves accuracy and helps convergence at higher values of $\chi$. But we would also like are calculation to be elegant, smooth, analytic, and intelligently designed. All  that is hard.
\end{itemize}

\subsection{Where do we stand?}
The rewiring method put forward by Nave and Levin\cite{LevinNave}  and by Gu and Wen\cite{GW} has a compelling elegance.  The replacement of four legged tensors by sums of three-legged ones is an excellent way of formulating the renormalization concept.  The next step, the evaluation of the three-legged tensors via SVD is attractive, but not equally compelling.  We follow many other authors in noting that this replacement depends only upon the local properties of the tensor being replaced, and not upon the global nature of the free energy calculation.  In contrast, the potential moving calculation contains a global optimization.    We might hope to combine the virtues of the two methods.

\section*{Appendix: Gauge freedom and invariance of the recursion step in isotropic TRG}\la{AGauge}
Below we prove that up to sign ambiguities, the renormalization step described in section \ref{subsec:svd} is invariant under rotations, i.e. if $\tilde{T}$ is obtained from $T$ by isotropic rotations, or component wise
\[
\tilde{T}_{ijkl}=\O_{ip}\O_{jq}\O_{kr}\O_{ls}T_{pqrs},
\]
where $\O$ is an orthogonal matrix,
then $\tilde{T}'=R(\tilde{T})=R(T)=T'$ up to sign ambiguities. Again component wise this reads
\[
\tilde{T}'_{ijkl}=D_{ip}D_{jq}D_{kr}D_{ls}T'_{pqrs},
\]
where $D_{ij}=\d_{ij}f_i$ where $f_i$ is either $+1$ or $-1$. This of course can be rectified by setting the sign of $D_{11}=+1$ and then making sure that $D_{1112},D_{1113}$ and $D_{1114}$ are all positive (provided that they do not vanish). Incorporating such a sign rectifying step into the TRG scheme results in
\[
\tilde{T}'_{ijkl}=T'_{ijkl}.
\]
An immediate corollary of this claim is that the isotropic response matix possess a null space whose dimension must be greater than that of the rotation group from which $\O$ was selected.
\subsection*{Proof}
We begin with considering the uniqueness of the SVD of a given, and its transformation under rotations. To avoid identity mis-interpretation we denote the $\chi^2$ valued index obtained from all the possible combinations of the $\chi$ valued indices $i$ and $j$ by $\{ij\}$. This makes its untangling simpler.

Let $T_{\{ij\}\{kl\}}$ be a diagonalizable matrix representing a rank four tensor and let
the SVD of $T$ be given by
\be
T_{\{ij\}\{kl\}}=U_{\{ij\}\a}\Lambda_{\a\b}V_{\{kl\}\b},
\label{eq:svd1}
\ee
where $\Lambda_{\a\b}=\lambda_a \d_{\a\b}$(no summation) are the principal values, and $U$ and $V$ are orthogonal matrices, then:
\begin{enumerate}
\item The columns of $U$ are the normalized eigenvectors of $TT^T$.
\item The rows of $V$ are the normalized eigenvectors of $T^TT$.
\item The orthogonal matrices are defined up to a sign: i.e. If $U,V$ are the orthogonal matrices obtained by some algorithm, and $\tilde{U},\tilde{V}$ orthogonal matrices obtained by a different yet equivalent algorithm then
\[
U^T\tilde{U}=V^T\tilde{V}=D,\qquad D_{ij}=\d_{ij}f_i,\quad f_i=\pm 1.
\]
\end{enumerate}
We now consider an orthogonal matrix which is the external product of two orthogonal matrices,  $\O_{\{ij\}\{kl\}}=O_{ik}O_{jl}$.
The rotation by $O$ of the rank four tensor $T_{ijkl}$ is equivalent to the rotation by $\O$ of the matrix  $T_{\{ij\}\{kl\}}$. By the above considerations the SVD of a rotated matrix, is the rotated SVD up to sign ambiguities.
If the SVD of $T$ is given by \eq{eq:svd1} then the SVD of the rotated matrix is given by
\[
\begin{aligned}
& \O _{\{ij\}\{mn\}}\O _{\{kl\}\{pq\}}T_{\{ij\}\{kl\}}\\
&=\O_{\{ij\}\{mn\}}U_{\{ij\}\a}D_{\a\gamma}\Lambda_{\gamma\d} D_{\b\d}V_{\{kl\}\b}\O_{\{kl\}\{pq\}},
\end{aligned}
\]
where again $D_{ij}=\d_{ij}f_i,\quad f_i=\pm 1$, accounts for the sign ambiguity. Setting $\sqrt{\Lambda}$ to be the nonnegative diagonal matrix whose square reproduces $\Lambda$, and using commutativity  and symmetry of products of diagonal matrices we have
\[
\begin{aligned}
\O_{\{ij\}\{mn\}}U_{\{ij\}\a}D_{\a\gamma}\sqrt{\Lambda}_{\gamma\d}=&
O_{im}O_{jn}U_{\{ij\}\a}\sqrt{\Lambda}_{\a\gamma}D_{\gamma\d}\\
=&O_{im}O_{jn}u_{ij\gamma}D_{\gamma\d}=\tilde{u}_{ij\d}
\end{aligned}
\]
As the last step in the renormalization includes a product of four such $u$ tensors, the rotations $O$ give unity and the only remnant is the sign ambiguity captured by $D$:
\[
\tilde{u}_{ij\a}\tilde{u}_{jk\b}\tilde{u}_{kl\gamma}\tilde{u}_{li\d}=
u_{ij\eta}u_{jk\mu}u_{kl\nu}u_{li\rho}D_{\a\eta}D_{\b\mu}D_{\gamma\nu}D_{\d\rho}.
\]

\section*{Acknowledgments}
We have had helpful conversations with Paul Wiegmann,  Ilya Gruzberg, Guifré Vidal, Ian Afflick,   and Brian Swingle.
This research was supported by the University of Chicago MRSEC, and NSF grant number DMR-0820054.
A.K. additionally acknowledges the support of St. Olaf College, and the Midstates Consortium.
E.E. additionally acknowledges the support of the Simons foundation.
and Z.W.  additionally acknowledges the support of the
University of Chicago's Office of the Vice President for Research and National
Laboratories and the University of Science and Technology of China.

\end{document}